\documentclass[preprint,12pt,sort&compress]{elsarticle}
\usepackage[a4paper,left=2.2cm,right=2.2cm,top=2.5cm,bottom=2.8cm]{geometry}
\usepackage{amsmath}
\usepackage{amsthm}
\usepackage{graphicx}
\usepackage{amsfonts}
\usepackage{color}
\newtheorem{defn}{Definition}
\newtheorem{thm}{Theorem}
\newtheorem{lem}{Lemma}
\newtheorem{prop}{Proposition}

\newtheorem{fact}{Fact}

\DeclareMathOperator*{\argmax}{arg\,max}
\DeclareMathOperator*{\argmin}{arg\,min}
\usepackage{algorithm}
\usepackage{algorithmic}

\usepackage{enumitem}
\usepackage{comment}

\journal{Journal of Computer and System Sciences}

\makeatletter
\def\ps@pprintTitle{%
  \let\@oddhead\@empty
  \let\@evenhead\@empty
  \def\@oddfoot{\reset@font\hfil\thepage\hfil}
  \let\@evenfoot\@oddfoot
}
\makeatother

\begin{document}

\begin{frontmatter}

\title{A Computational Analysis of Strategic Nominations: Modeling Equilibrium and Complexity in Organizational Elections\tnoteref{label_title}}
\tnotetext[label_title]{A previous version appears at the 6th Games, Agents, and Incentives Workshop (GAIW-24)~\cite{LLC2024}. Held as part of the Workshops at the 22nd International Conference on Autonomous Agents and Multiagent Systems.}

\author[CCLin]{Chuang-Chieh Lin}
\ead{josephcclin@mail.ntou.edu.tw}

\author[CJLu]{Chi-Jen Lu}
\ead{cjlu@iis.sinica.edu.tw}

\author[PAChen]{Po-An Chen\corref{cor}}
\ead{poanchen@nycu.edu.tw}

\author[CCHung]{Chih-Chieh Hung\corref{cor}}
\ead{smalloshin@nchu.edu.tw}

\affiliation[CCLin]{organization={Department of Computer Science and Engineering, National Taiwan Ocean University},
    addressline={No.2, Beining Rd., Jhongjheng Dist.}, 
    city={Keelung City},
    postcode={202301}, 
    country={Taiwan}}

\affiliation[CJLu]{organization={Institute of Information Science, Academia Sinica},
    addressline={128 Academia Rd. Sect. 2, Nankang Dist.}, 
    city={Taipei},
    postcode={11529}, 
    country={Taiwan}}

\affiliation[PAChen]{organization={Institute of Information Management, National Yang Ming Chiao Tung University},
            addressline={1001 University Rd., East Dist.}, 
            city={Hsinchu City},
            postcode={300}, 
            country={Taiwan}}

\affiliation[CCHung]{organization={Department of Management Information Systems, National Chung Hsing University},
            addressline={145 Xingda Rd., South Dist.}, 
            city={Taichung City},
            postcode={402202}, 
            country={Taiwan}}
\cortext[cor]{Corresponding authors}

\begin{abstract}
We study organizational elections in which each group nominates one candidate and receives as payoff its members’ expected utility under a probabilistic winning rule. We empirically justify a standard monotonicity assumption by simulating two- and three-group elections, finding that a candidate’s aggregate voter utility correlates monotonically with win probability. For three or more groups, we show that pure-strategy Nash equilibria (PSNE) may fail to exist even under egoistic preferences, and that deciding PSNE existence is {\sf NP}-complete in a succinct (general form) representation. For cross-monotone winning-probability functions, we give simple sufficient conditions for PSNE existence and an FPT algorithm to compute one, parameterized by the number of irresolute groups and nominating depth. Finally, for cross-monotone, order-preserving winning-probability functions, we bound the price of anarchy of egoistic games by the number of groups.
\end{abstract}


\end{frontmatter}

\section{Introduction}
\label{sec:intro}

\subsection{Background and Motivation}
\label{subsec:background}

Political competition is often mediated by parties nominating candidates and competing in elections.
Duverger's law suggests that plurality voting tends to favor two-party systems~\cite{Duv54}, and
Dellis~\cite{Del2013} analyzed conditions under which plurality and related ballot restrictions lead to two-party outcomes.
Motivated by these observations, Lin et al.~\cite{LLC2021} proposed a macro-level non-cooperative model, the \emph{two-party election game}, bypassing micro-level voting details.

In the two-party election game, each party (group) is a strategic player whose pure strategies are its candidates.
Given a strategy profile, the winning-probability (WP) function determines which candidate wins (possibly probabilistically),
and a group's payoff is the expected utility obtained by its members.
Lin et al.~\cite{LLC2021} introduced two structural assumptions—\emph{monotonicity} and \emph{egoism} (Definitions~\ref{def:monotone} and~\ref{defn:egoistic})—and proved existence of pure-strategy Nash equilibria (PSNE) and constant price-of-anarchy (PoA) bounds for two parties under linear and softmax WP rules.

Beyond political elections, the same framework captures organizational competitions in which multiple groups nominate representatives. 
For example, consider an organization, such as a large corporation, divided into multiple factions or groups. Each group (e.g., Front-end Engineers, Back-end Engineers, QA/Testing Engineers, etc.) would prefer a candidate from their own group because that candidate would best advance their workflows and goals. As another example, consider that a computer science department, which encompasses groups of Theory \& Algorithms, Systems \& Architecture, Artificial Intelligence \& Machine Learning, etc., is hiring a new professor. The ``Theory \& Algorithms'' faculty strongly prefer a candidate whose expertise in algorithm design and analysis aligns with their research interests and can provide relevant collaboration and support. 
Analogous to political elections, a nominated candidate may benefit both its own members and other groups' members. 
Throughout this paper, we use \emph{members} to denote individuals and \emph{groups} to denote strategic entities; political parties are a special case of groups.

PSNE are particularly desirable as they provide stable outcomes under unilateral deviations, yet they need not exist in finite games~\cite{OR94}.
While Lin et al.~\cite{LLC2021} establishes strong stability and efficiency guarantees in the two-party case under monotonicity and egoism,
it remains unclear how these guarantees extend to elections with three or more competing groups.
This motivates our study of equilibrium existence, computational complexity, algorithmic tractability, and inefficiency (PoA) in multi-group egoistic election games.

\subsection{Our Contributions}
\label{subsec:contribution}

In this work, we generalize Lin \textit{et al}.'s work~\cite{LLC2021} to deal with $m\geq 2$ groups. 
Throughout this paper, we consider the monotone egoistic election game and omit the adjective ``monotone'' when the context is clear. 
For our algorithmic result (Sect.~\ref{sec:fpt-algo}), we assume that the WP function is \emph{cross-monotone} (Definition~\ref{def:crossmono}), which is automatically satisfied in the two-party case ($m=2$). 
For our PoA analysis (Sect.~\ref{sec:PoA}), we additionally assume the WP function is \emph{order-preserving} (Definition~\ref{def:orderpreserving}).
For the egoistic election game of two or more groups, we aim at investigating the following questions:
\begin{enumerate}
    \item Can we have any explanation of the monotone property introduced and assumed in~\cite{LLC2021}? Does the property still make sense in the competition of more than two groups or organizations? 
    \item Does the egoistic election game using the softmax function to calculate the winning probability of a candidate against its opponents always admit a PSNE, even for three or more groups? Does the function for computing such a winning probability matter? 
    \item What is the computational complexity of computing a PSNE of the egoistic election game when two or more groups are involved in general? 
    \item How large is the price of anarchy of the egoistic election game of two or more groups? Is it still bounded by a constant?
\end{enumerate}
    
\paragraph{Our answers to the above questions are summarized as follows} 

\begin{enumerate}
    \item We conduct experiments to simulate the voting outcomes in the competitions between two and three groups to exemplify how the aggregation of micro behaviors leads to the monotone property in the macro perspective. 
    \item We give examples to confirm that a PSNE does not always exist in the egoistic election game of three or more groups even using the softmax function to calculate the winning probability of a candidate. 
    Nevertheless, we propose two sufficient conditions for the egoistic election game to have a PSNE. 
    \item We prove that to determine if a PSNE exists in the egoistic election game in the general-form representation is {\sf NP}-complete. Moreover, when the WP function is cross-monotone, based on the two sufficient conditions, we identify two parameters 
    of the egoistic election game, and propose a fixed-parameter tractable algorithm to find a PSNE of the game if it exists. Namely, under cross-monotonicity, a PSNE of the egoistic election game can be found in time polynomial in the number of groups and number of candidates in each group if the two proposed parameters are as small as constants. 
    \item Perhaps surprisingly, assuming the WP function is cross-monotone and order-preserving, we prove that the PoA of the egoistic election game is at most
    the number of competing groups, i.e., $m$. This bound is essentially tight: under the hardmax WP rule, we construct instances whose PoA attains~$m$.
    In particular, when $m=2$, our upper bound gives the PoA is at most~$2$. Combined with the known the lower bound $2$ for the two-party election game in~\cite{LLC2021} (using the hardmax or the softmax WP function), this yields tightness for the two-party case.
    Finally, we show that the order-preserving assumption is necessary in general by giving an egoistic instance with unbounded PoA under a cross-monotone but 
    non-order-preserving WP function.
\end{enumerate}

\subsection{Related Work}
\label{subsec:related_work}

\paragraph{On Duverger's law} Duverger's law suggests plurality voting in favor of the two-party system~\cite{Duv54}, and this can be explained either by the strategic 
behavior of the voters~\cite{Del2013,Fed92,Fey97,MW93,Pal89} or that of the 
candidates~\cite{Cal05,CW07,Pal84,Web92}. The latter considers models where two selected candidates face the third candidate as a potential threat. Dellis~\cite{Del2013} explained (with mild assumptions on voters'
preferences) why a two-party system emerges under plurality voting and other voting procedures permitting truncated ballots. Nevertheless, why the two candidates are selected is not discussed.

\paragraph{On spatial theory of voting} Most of the works on equilibria of a political competition are mainly based on {\em Spatial Theory of Voting}~\cite{Hot29,Dow57,Pal84,Web92}, 
which can be traced back to~\cite{Hot29}. 
In such settings, there are two parties and voters with single-peaked preferences over a unidimensional metric space. Each party 
chooses a kind of ``policy'' that is as close as possible to voters' preferences. When the policy space is unidimensional, the Spatial Theory of Voting states that the parties' strategies would be determined by the median voter's preference. However, pure-strategy Nash equilibria may not exist for policies over a multi-dimensional space~\cite{Dug16}.

\paragraph{On the Hotelling-Downs model} 
Hotelling-Downs model~\cite{Hot29} originally considers the problem that two strategic ice cream vendors along a stretch of beach try to attract as many customers as possible by placing themselves. This framework can be extended to the setting that two parties nominate their candidates on a political spectrum. The model has variations involving \emph{multiple} agents with \emph{restricted options}, and 
this is in line with the competition of multiple parties with 
a few nominees as possible candidates. For the variation of Hotelling-Downs model as such, Harrenstein et al.~\cite{HLST21} show that computing a Nash equilibrium is 
{\sf NP}-complete in general but can be done in linear time when there are only two competing parties. Sabato et al.~\cite{SORR17} consider \emph{real candidacy games}, in which competing agents select their positions from their corresponding intervals on the real line and then the outcome of the competition follows a given social choice rule. They establish conditions for the existence of a PSNE, yet the computational complexity is not discussed. For such Hotelling-Downs-like models, players only care about winning, while in our work, group players aim at maximizing the expected utility of their members.

\paragraph{Other work modeling an election as a game} Ding and Lin~\cite{DL2014} considered open list proportional representation as the election mechanism which has been used in European elections for parliament seats. Each voter is given a set of lists of candidates to vote for and exactly one list will be cast. The mechanism proceeds in two rounds to compute the winners. They formulate the election of exactly two parties as a two-player zero-sum game and show that the game always admits a PSNE while it is {\sf NP}-hard to compute it. 
The setting in Laslier's work~\cite{L00a,L00b} is close to our work. Two parties are considered as two players, each of which provides a finite set of alternatives for the voters. Yet, a party's strategy is viewed as a ``mixed one'' instead, that is, a non-spatial alternative is identified by a fraction of the voters. With standard analysis, the mixed-strategy Nash equilibrium is guaranteed to exist. Compared with our work, the winning probabilities, the expected utility of each group as well as the price of anarchy are not considered in~\cite{L00a,L00b}.

Very recently, Cechl\'{a}rov\'{a} et al.~\cite{cechlarova2023hardness} consider  candidate nomination with parties under classical voting rules, which asks whether a designated candidate can or must (corresponding to a possible or necessary president, resp.) win when each party nominates a single candidate and voters’ full preferences are known. They adopt a unique-winner model and show {\sf NP}-completeness and {\sf coNP}-completeness for these questions, respectively, for $k$-approval/$k$-veto, plurality with runoff, maximin, Copeland, and Llull, etc. They also give integer programming formulations as well as computational experiments. In contrast, our work views parties as strategic players in an \emph{egoistic} election game under a monotone winning-probability (WP) function: we characterize regimes that guarantee PSNE or not, show that deciding PSNE existence is {\sf NP}-complete, and---under the additional cross-monotonicity property---design a fixed-parameter algorithm for PSNE computation; moreover, we bound the price of anarchy by the number of competing groups.

\subsection{Discussion of Our Work and the Organization}
\label{subsec:discussion_organization}

 \paragraph{Discussion of our work} 
Most of the literature on the voting theory and the game of elections focuses on the voters' behavior at a \emph{micro}-level and outcomes by various election rules, which proceed with either one or multiple rounds and aim to bring one or multiple winners. Voters can be strategic and have different preferences for the candidates. Their behaviors can even be affected by the other voters (e.g., see~\cite{AFS19,BFM18}) and also dependent on the social choice rule, such as the design of ballots, rounds of selection, and the allocation rule, etc. In this work, we bypass the above-involved factors in an election and focus on a \emph{macro-level} analysis instead. 
The discussions can be applied to organizational elections in which the social choice rule can be different from that in political elections. 
By introducing uncertainty in the competition between candidates participating in the competition, the payoff as the expected utility for the members of a group can also be regarded as the sum of fractional social welfare one group's members receive from all the competing candidates. 
Moreover, the monotonicity of the game is arguably natural in the sense that a group can attract more organization members when it nominates a candidate who benefits the members more in the organization. When the WP function is both cross-monotone and order-preserving, we show that the PoA of the game is upper bounded by the number of competing groups~$m$. Without the order-preserving guarantee, we also provide an example which has unbounded PoA. Our work provides an alternative perspective and simpler evidence on the inefficiency of multiple-group competition, which complements previous relevant work.

\paragraph{What is new beyond the workshop version (GAIW'24)}
This journal version substantially extends our preliminary workshop paper~\cite{LLC2024} presented in GAIW'24. 
\begin{enumerate}[label = (\roman*)]
\item \emph{Comprehensive Framing and Motivation}: This work has been fundamentally reframed from a specific ``political party" context to a more general and broadly applicable model of ``organizational elections". More discussions and expanded related work are provided to motivate this broader scope. 
\item \emph{Experimental Justification of the Setting}: We have conducted experiments to validate the assumption in our model on the monotone property of the winning probability. 
\item \emph{Full Integration of Technical Results}: This work is  complete and self-contained. All definitions and proofs, including the detailed {\sf NP}-completeness reduction, are now fully integrated and elaborated. 
\end{enumerate}

\paragraph{Organization of this paper} 
Sect.~\ref{sec:preliminaries} introduces preliminaries and notation. 
Experimental results for illustrating the monotone property are shown in Sect.~\ref{sec:monotone_verification}. 
In Sect.~\ref{sec:hardness_no_PSNE_examples}, we use examples to contrast different WP rules:
under hardmax the egoistic election game always admits a PSNE, whereas under softmax there exist three-group instances with no PSNE.
We also establish that deciding whether a given egoistic election game admits a PSNE is {\sf NP}-complete.
In Sect.~\ref{sec:fpt-algo}, 
we propose two sufficient conditions for the egoistic election game with a cross-monotone WP function to admit a PSNE. Then, we propose a fixed-parameter tractable algorithm to compute a PSNE of the game in the general form representation. 
Finally, Sect.~\ref{sec:PoA} proves PoA upper bounds for cross-monotone and order-preserving WP functions.
Sect.~\ref{sec:future} concludes with directions for future work.


\section{Preliminaries}
\label{sec:preliminaries}


For an integer $k>0$, let $[k]$ denote the set $\{1,2,\ldots,k\}$. We assume that the organization consists of members, and each member belongs to one of the $m\geq 2$ groups $\mathcal{P}_1,\mathcal{P}_2,\ldots,\mathcal{P}_m$. These $m$ groups compete in an election competition. Each group $\mathcal{P}_i$, having $n_i\geq 2$ candidates $x_{i,1}, x_{i,2}, \ldots, x_{i,n_i}$ for each $i\in [m]$, has to designate one candidate to participate in the competition. Let $n = \max_{i\in [m]} n_i$ denote the maximum number of candidates in a group. Let $u_j(x_{i,s})$ denote the (aggregate) \emph{utility} of all members in group $\mathcal{P}_j$ when candidate $x_{i,s}$ of group $\mathcal{P}_i$ is elected, for $i,j\in [m], s\in[n_{i}]$. 
Since $u_j(x_{i,s})$ is defined as an aggregate over all members in~$\mathcal{P}_j$, group sizes are implicitly accounted for in $u(x_{i,s})$.
For each $i\in [m]$ and $s\in [n_i]$, we define the \emph{social utility} of  candidate $x_{i,s}$ as $u(x_{i,s}) := \sum_{j\in [m]}u_j(x_{i,s})$ which represents the aggregate utility that $x_{i,s}$ brings to all members of the organization. Assume that the social utility is nonnegative and bounded, specifically, we assume $u(x_{i,s})\in [0, \beta]$ for some real $\beta\geq 1$, for each $i\in [m], s\in [n_i]$.
Assume that candidates in each group are 
sorted according to the utility for its group's members. 
Namely, we assume that $u_i(x_{i,1})\geq u_i(x_{i,2})\geq \ldots \geq u_i(x_{i,n_i})$ for each $i\in [m]$. 

We focus on the \emph{election game}, which models the election competition as a game of $m$ strategic \emph{group players}. With a slight abuse of notation, $\mathcal{P}_i$ also denotes the group player with respect to group $\mathcal{P}_i$. Each group player $\mathcal{P}_i$, $i\in [m]$, has $n_i$ \emph{pure strategies}, each of which is a candidate selected to participate in the competition. Suppose that $\mathcal{P}_i$ designates candidates $x_{i,s_i}$ for $i\in [m]$ and let $\mathbf{s} = (x_{1,s_1},x_{2,s_2},\ldots,x_{m,s_m})$ (or simply $(s_1,s_2,\ldots,s_m)$ when it is clear from the context) be the \emph{profile} of the designated candidates of all the group players. We denote by $\mathbf{s}_{-i}$ the profile $\mathbf{s}$ except group player $\mathcal{P}_i$'s strategy~$s_i$. Let $p_{i,\mathbf{s}}$ denote the probability of group player~$\mathcal{P}_i$ winning the competition with respect to profile~$\mathbf{s}$. We consider an assumption as a desired \emph{monotone property} that \emph{a group wins the competition with higher or equal odds if it selects a candidate with a higher social utility}.
The odds of winning depend on the social utility brought by the candidates. 
\begin{defn}[Monotone Property]\label{def:monotone}
The election game is \emph{monotone} if for every group $\mathcal{P}_i$, $i\in[m]$, every fixed $\mathbf{s}_{-i}$,
and any two strategies $s_i,s_i'\in [n_i]$ of~$\mathcal{P}_i$, whenever
\[
u(x_{i,s_i'}) \geq u(x_{i,s_i}),
\]
we have
\[
p_{i,(s_i',\mathbf{s}_{-i})} \geq p_{i,(s_i,\mathbf{s}_{-i})}.
\]
\end{defn}
The payoff of group player $\mathcal{P}_i$ given the profile $\mathbf{s}$ is denoted by $r_{i,\mathbf{s}}$, which is the \emph{expected utility} that group~$\mathcal{P}_i$'s members obtain in~$\mathbf{s}$. Namely, 
$r_{i}(\mathbf{s}) = \sum_{j\in [m]} p_{j,\mathbf{s}}u_i(x_{j,s_j})$,
which can be computed in~$O(m)$ time for each~$i$. 
\paragraph{Remark} We regard a group in an organization, which consists of its members, as a collective concept. A candidate is nominated by the members of the group to compete in an election competition. Members of a group can gain utility not only from the candidate nominated by their group but also from anyone by other competing groups. Since the winning candidate serves for the whole organization, that is, for \emph{all} the organization members, 
by considering the payoff as the expected utility we can formulate the payoff of a group player no matter it wins or loses.

As defined and discussed in~\cite{LLC2021}, we can concretely formulate $p_{i,\mathbf{s}}$ and preserve the monotone property as the following examples.
\begin{itemize}
\item The hardmax function: 
\[
p_{i,\mathbf{s}} = \left\{\begin{array}{ll}
1 & \mbox{ if } i = \min\{\argmax_{j\in [m]} u(x_{j,s_j})\}\\
0 & \mbox{ otherwise.}
\end{array}\right.
\]
    \begin{itemize}
        \item The hardmax function simply allocates all probability mass to the candidate of maximum social utility. If the size of
        $\argmax_{j\in [m]} u(x_{j,s_j})$ is larger than~1, then all the probability mass is allocated to the one with minimum index in~$\argmax_{j\in [m]} u(x_{j,s_j})$. 
    \end{itemize}
\item The softmax function~\cite{Kul00,SB98}: 
	\[p_{i,\mathbf{s}} := \frac{e^{u(x_{i,s_i})/\beta}}{\sum_{j\in [m]} e^{u(x_{j,s_j})/\beta}}.\]
    \begin{itemize}
	    \item The softmax function 
        is formulated as the ratio of one exponential normalized social utility to the sum over all nominated candidates. Clearly, it is \emph{nonlinear} in the social utility $u(x_{i,s_i})$  and produces a probability strictly in $(0, 1)$.
	\end{itemize}
 \vspace{5pt}
\item The natural function~\cite{Bra54,YBKJ2012,DWH2020}: \[p_{i,\mathbf{s}} := \frac{u(x_{i,s_i})}{\sum_{j\in [m]} u(x_{j,s_j})}.\] 
    \begin{itemize}
    \item We treat the probability $p_{i,\mathbf{s}}$ as ratio of the social utility brought by a candidate to the sum of the social utility brought by all candidates.\footnote{We assume that $\sum_{j\in [m]} u(x_{j,s_j})>0$ for the natural function.} This function is \emph{linear} in the social utility $u(x_{i,s_i})$ and produces a probability in~$[0, 1]$
    \end{itemize}
\end{itemize}

We call the functions calculating the winning probability of a candidate against the others \emph{WP functions}. The WP functions by which the egoistic election game is monotone are called \emph{monotone WP functions}. The hardmax function is monotone since raising the social utility never makes a group player lose. By Fact~\ref{lem:fractional_monotone} below, we know that the softmax 
function is also a monotone WP function. Throughout this paper, we consider monotone WP functions unless otherwise specified. 
\begin{fact}\label{lem:fractional_monotone}
Let $a, b > 0$ be two positive real numbers such that $a<b$. Then, for any $d>0$, $a/b < (a+d)/(b+d)$. 
\end{fact}
\begin{proof}
Let $a, b > 0$ be two positive real numbers such that $a<b$. Then $a/b - (a+d)/(b+d) = d(a-b)/(b^2+bd)$. Since $d>0$, $a<b$, and $b(b+d)>0$, we have $a/b - (a+d)/(b+d) < 0$. 
\end{proof}

\paragraph{Cross-monotonicity}
For several subsequent arguments (in particular, Lemma~\ref{lem:dominated_Not_PSNE} and its applications),
we will use a slightly stronger property than standard monotonicity, capturing that probability mass
shifted toward one group should come at the expense of the others. We define it below.
\begin{defn}[Cross-monotone WP function]\label{def:crossmono}
A monotone WP function is \emph{cross-monotone} if for every group $i\in[m]$,
every fixed $\mathbf{s}_{-i}$, and any two strategies $s_i,s_i'\in[n_i]$, whenever
\[
u(x_{i,s_i'}) \geq u(x_{i,s_i}) \quad \text{and} \quad p_{i,(s_i',\mathbf{s}_{-i})} > p_{i,(s_i,\mathbf{s}_{-i})},
\]
we have
\[
p_{j,(s_i',\mathbf{s}_{-i})} \leq p_{j,(s_i,\mathbf{s}_{-i})} \quad \text{for all } j\neq i.
\]
\end{defn}

\begin{defn}[Order-preserving WP function]\label{def:orderpreserving}
A monotone WP function is \emph{order-preserving} if for every strategy profile $\mathbf{s}$ and any two groups
$i,j\in[m]$, $u(x_{i,s_i}) > u(x_{j,s_j})$ implies that $p_{i,\mathbf{s}} \ge p_{j,\mathbf{s}}$.
\end{defn}

\noindent\textbf{Remark.} When $m=2$, cross-monotonicity holds automatically for any WP function,
since $p_{1,\mathbf{s}}+p_{2,\mathbf{s}}=1$ for all $\mathbf{s}$. In contrast, the order-preserving property 
is not automatic even in the two-party case.
It additionally requires that the candidate with higher social utility is at least as likely to win
(i.e., $u(x_{1,s_1})>u(x_{2,s_2})$ implies $p_{1,\mathbf{s}}\geq 1/2$).
Nevertheless, this property is satisfied by common WP functions such as hardmax, softmax, and the natural WP functions.

\begin{lem}\label{lem:wp-crossmono}
The hardmax, softmax, and natural WP functions are cross-monotone and order-preserving.
\end{lem}
\begin{proof}
First, let us prove the cross-monotone property. 
For the natural and softmax WP functions, we can write
\begin{equation}\label{eqn:proof_natural_softmax}
p_{i,\mathbf{s}}=\frac{w(u(x_{i,s_i}))}{\sum_{k\in[m]} w(u(x_{k,s_k}))},
\end{equation}
where $w(z)=z$ for the natural WP function and $w(z)=e^{z/\beta}$ for the softmax WP function, and $w(\cdot)$ is positive and increasing.
Fix $\mathbf{s}_{-i}$. If $u(x_{i,\mathbf{s}_i'})\geq u(x_{i,\mathbf{s}_i})$, then the denominator weakly increases while
the numerator for any $j\neq i$ stays unchanged, so $p_{j,(s_i',\mathbf{s}_{-i})}\leq p_{j,(s_i,\mathbf{s}_{-i})}$.
For the hardmax WP function, increasing $u(x_{i,s_i})$ can only make $\mathcal{P}_i$ enter (or improve within)
the argmax set, which cannot increase the winning probability of any other group.

Second, we prove the order-preserving property. 
Fix a profile $\mathbf{s}$ and two groups $i,j\in[m]$ such that $u(x_{i,s_i}) > u(x_{j,s_j})$. 
Under the hardmax WP function, group $\mathcal{P}_j$ can have $p_{j,\mathbf{s}}=1$ 
only if it is selected as the (tie-broken) maximizer of $u(x_{\ell,s_\ell})$.
However, $u(x_{i,s_i})>u(x_{j,s_j})$ implies that $j$ is not a maximizer, hence $p_{j,\mathbf{s}}=0$, and therefore
$p_{i,\mathbf{s}}\geq 0 = p_{j,\mathbf{s}}$.
As for the softmax and natural WP functions, both of them have the form as Equation~\ref{eqn:proof_natural_softmax}. 
Clearly, $u(x_{i,s_i})>u(x_{j,s_j})$ implies 
$w(u(x_{i,s_i}))>w(u(x_{j,s_j}))$ and hence $p_{i,\mathbf{s}}>p_{j,\mathbf{s}}$.
\end{proof}

We define the {\em social welfare} of the profile $\mathbf{s}$ as 
\begin{align*}
    SW(\mathbf{s}) = \sum_{i\in [m]} r_{i}(\mathbf{s}) &= \sum_{i\in [m]}\sum_{j\in [m]} p_{j,\mathbf{s}}u_i(x_{j,s_j})\\
    &= \sum_{j\in[m]}p_{j,\mathbf{s}} \sum_{i\in [m]} u_i(x_{j,s_j})\\ 
    &= \sum_{j\in[m]}p_{j,\mathbf{s}} u(x_{j,s_j}).
\end{align*} We say that 
a profile $\mathbf{s}$ is a {\em pure-strategy Nash equilibrium} (PSNE) if $r_{i}(s'_{i},\mathbf{s}_{-i})\leq r_{i}(\mathbf{s})$ for any~$s'_i$ with $\mathbf{s}_{-i}$ fixed. That is, in $\mathbf{s}$, none of the group players has the incentive to deviate from its current strategy unilaterally. The {\em (pure) price of anarchy} (PoA) of the game~$\mathcal{G}$ is defined as 
\[
\mbox{PoA}(\mathcal{G})=\frac{SW(\mathbf{s}^*)}{SW(\hat{\mathbf{s}})} = 
\frac{\sum_{j\in [m]} r_{j}(\mathbf{s}^*)}{\sum_{j\in [m]} r_{j}(\hat{\mathbf{s}})},
\]
where $\mathbf{s}^* = \argmax_{\mathbf{s}\in\prod_{i\in [m]}[n_i]} SW(\mathbf{s})$ is the {\em optimal profile}, 
which has the highest social welfare among all possible profiles, and $$\hat{\mathbf{s}} = \argmin_{\mathbf{s}\in\prod_{i\in [m]} [n_i], \,\mathbf{s}\mbox{\scriptsize \;is a PSNE}} SW(\mathbf{s})$$
is the PSNE with the lowest social welfare\footnote{For $\hat{\mathbf{s}}$ 
being the PSNE with the best (i.e., highest) social welfare, the \emph{(pure) price of stability} (PoS) of~$\mathcal{G}$ is defined as~$SW(\mathbf{s}^*)/SW(\hat{\mathbf{s}})$.}. 
Note that an egoistic election game may not necessarily admit a PSNE so an upper bound on the PoA is defined over only games with PSNE, i.e., $\max_{\mathcal{G}\mbox{\scriptsize\, has a PSNE}}\mbox{PoA}(\mathcal{G})$, accordingly. 

We will use the following \emph{egoistic property} throughout this paper. 

\begin{defn}[Egoistic Property~\cite{LLC2021}]\label{defn:egoistic}
We call the election game \emph{egoistic} if for every $i\in[m]$ and every $j\in[m]\setminus\{i\}$,
and for all $s_i\in[n_i]$ and $s_j\in[n_j]$, it holds that
\[
u_i(x_{i,s_i}) > u_i(x_{j,s_j}).
\]
\end{defn}
That is, each group strictly prefers any of its own candidates to any candidate nominated by another group.

\begin{defn}[Surpass \& Weakly Surpass]
\label{defn:strategy_domination}
Given a profile $\mathbf{s}$, for $i\in [m]$, 
Strategy $x_{i,s_i}$ {\em weakly surpasses} $x_{i,s_i'}$ if $s_i < s_i'$ (i.e., $u_i(x_{i,s_i})\geq u_i(x_{i,s_i'})$) and 
$u(x_{i,s_i})\geq u(x_{i,s_i'})$. 
Strategy $x_{i,s_i}$ {\em surpasses} $x_{i,s_i'}$ if $x_{i,s_i}$ weakly surpasses $x_{i,s_i'}$ and either $p_{i,\mathbf{s}} > p_{i,(s_i',\mathbf{s}_{-i})}$ or $u_i(x_{i,s})>u_i(x_{i,s'})$ and $p_{i,\mathbf{s}}>0$. 
\end{defn}
By Definition~\ref{defn:strategy_domination}, a candidate surpasses another one in the same group if it brings more (resp., no less) utility to its members and has no lower (resp., higher) winning probability. 

\section{Illustrative Simulation for the Monotonicity Assumption}
\label{sec:monotone_verification}

This section provides an illustrative simulation to support the monotonicity assumption. 
It does not introduce new model assumptions or theorems, but offers intuition.
Its purpose is to show that several simple micro-level voting heuristics induce a winning-probability curve
that is monotone in a candidate's (normalized) aggregate advantage, which supports the assumption
used in~\cite{LLC2021} and in this work.
Recall from Sect.~2 that the social utility of a candidate $x$ is
$u(x)=\sum_{j\in[m]} u_j(x)$; in the simulation we use a scalar ``budget'' as a convenient proxy for $u(x)$.

Let us consider two groups $A$ and~$B$ with budgets $\text{budget}_A,\text{budget}_B\in[0,1]$.
We generate voter-level utilities by randomly splitting each budget among $N=100$ voters as 
\begin{align*}
& (\mu_1,\mu_2,\ldots,\mu_N) \sim \text{budget}_A \cdot \text{Dirichlet}(1,1,\ldots,1),\\
& (\mu'_1,\mu'_2,\ldots,\mu'_N) \sim \text{budget}_B \cdot \text{Dirichlet}(1,1,\ldots,1),
\end{align*}
so that $\sum_{i=1}^N \mu_i=\text{budget}_A$ and $\sum_{i=1}^N \mu'_i=\text{budget}_B$. 
Here $\text{Dirichlet}(1,1,\ldots,1)$ is uniform over the $N$-simplex, and scaling by $\text{budget}_A$
ensures the allocated utilities sum to $\text{budget}_A$.
Given $(\mu_i,\mu'_i)$, voter $v_i$ votes for $A$ with probability $q_i$ defined by one of the following heuristics:
\begin{itemize}
\item Linear:\quad $\displaystyle{q_i= \frac{1}{2} + \frac{\mu_i-\mu'_i}{2}}$.
\item Natural:\quad $q_i=\dfrac{\mu_i+\epsilon}{\mu_i+\mu'_i+2\epsilon}$.
\item Softmax:\quad $q_i=\dfrac{e^{\mu_i}}{e^{\mu_i}+e^{\mu'_i}}$.
\item Best-Response\footnote{This best-response rule is a voter-level heuristic used only in the simulation. It should not be confused with the hardmax WP function in Sect.~\ref{sec:preliminaries}, which is defined on candidates' social utilities.}:\quad $q_i=
\begin{cases}
1 & \text{if }\mu_i>\mu'_i,\\
0 & \text{if }\mu_i<\mu'_i,\\
1/2 & \text{if }\mu_i=\mu'_i,
\end{cases}$
(i.e., ties are broken uniformly).
\end{itemize}
Here $\epsilon=10^{-9}$ prevents division by zero in the natural rule.
We sample each vote independently as $\text{Bernoulli}(q_i)$ and declare the winner by majority vote;
ties are broken by a fair coin. For each budget pair, we repeat the election $R=1{,}000$ times to estimate $A$'s 
winning probability and a $95\%$ confidence interval.

\begin{figure}[ht]
  \centering
  \includegraphics[width=0.95\linewidth]{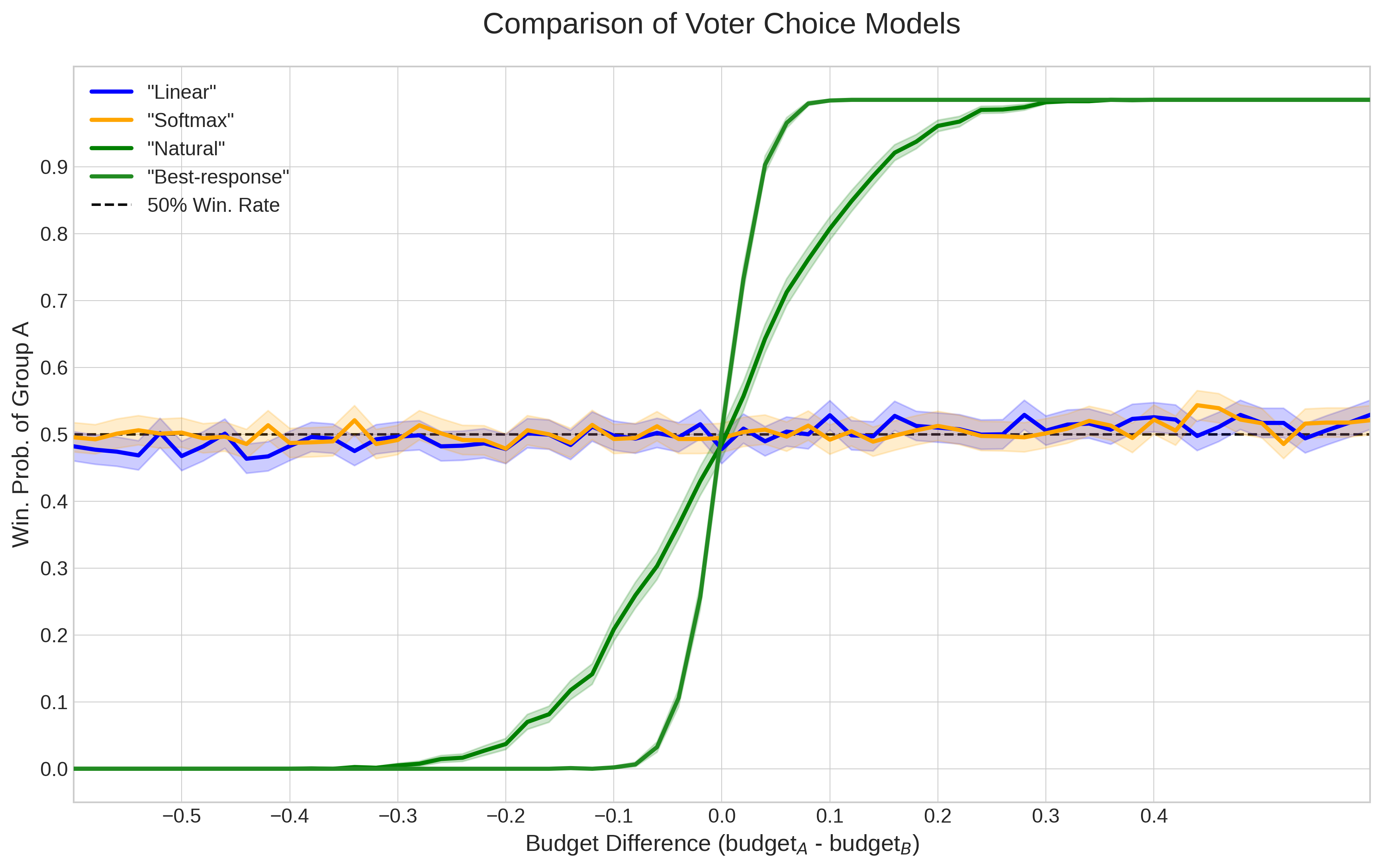}
  \caption{Two-group simulation. The shaded band indicates a $95\%$ confidence interval.}
  \label{fig:voting_2groups}
\end{figure}

For visualization, we fix $\text{budget}_B=0.5$ and vary $\text{budget}_A\in[0,1]$.
Fig.~\ref{fig:voting_2groups} shows that $A$'s estimated winning probability is nondecreasing as $\text{budget}_A$ increases.
Moreover, Spearman's rank correlation between $\text{budget}_A-\text{budget}_B$ and the estimated winning probability
is $0.9069$ (linear) and $0.7055$ (softmax), with $p$-values $<0.001$.
This provides a simplified micro-level illustrating our monotonicity assumption on the WP function in Sect.~\ref{sec:preliminaries}.

\begin{figure}[ht]
  \centering
  \includegraphics[width=0.95\linewidth]{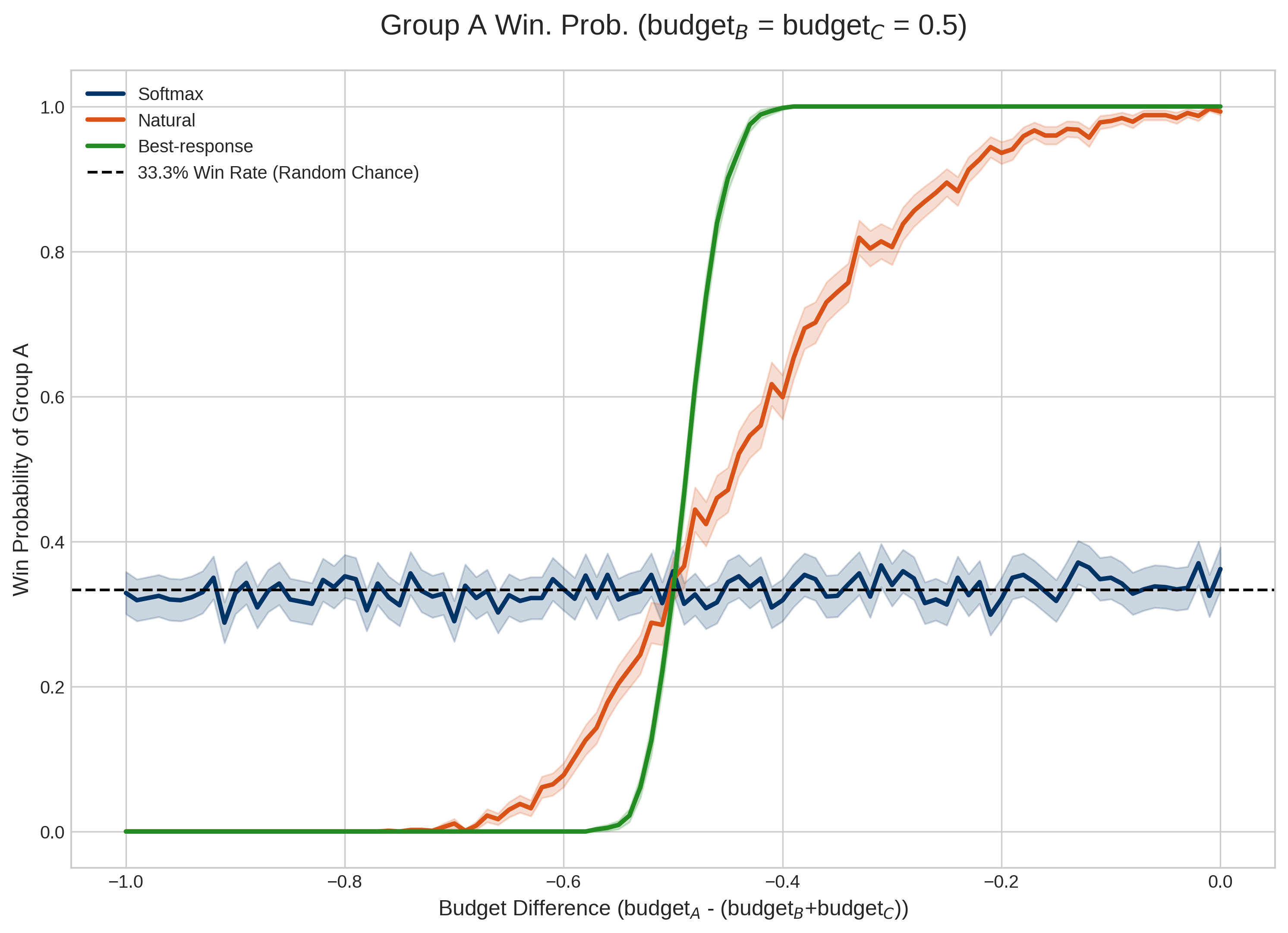}
  \caption{Three-group simulation. The shaded band indicates a $95\%$ confidence interval.}
  \label{fig:voting_3groups}
\end{figure}

We also consider three groups $A,B,C$ with budgets in $[0,1]$ and voter utilities
$(\mu_i,\mu'_i,\mu''_i)$ generated analogously by independent Dirichlet splits.
For each voter $v_i$, the probability of voting for $A$ is defined by:
\begin{itemize}
\item Natural:\quad $q_i^A=\dfrac{\mu_i+\epsilon}{\mu_i+\mu'_i+\mu''_i+3\epsilon}$.
\item Softmax:\quad $q_i^A=\dfrac{e^{\mu_i}}{e^{\mu_i}+e^{\mu'_i}+e^{\mu''_i}}$.
\item Best-Response:\quad $q_i^A = 
\begin{cases}
1   & \text{if }\mu_i > \max\{\mu_i',\mu_i''\},\\[2pt]
0   & \text{if }\mu_i < \max\{\mu_i',\mu_i''\},\\[2pt]
1/2 & \text{if }\mu_i=\mu_i' > \mu_i'' \text{ or } \mu_i=\mu_i'' > \mu_i',\\[2pt]
1/3 & \text{if }\mu_i=\mu_i'=\mu_i'',
\end{cases}
\;\; \text{(ties are broken uniformly).}$
\end{itemize}
We again determine the winner by plurality/majority aggregation (as implemented in the simulation) and repeat $R=1{,}000$ trials.
For visualization, we fix $\text{budget}_B=\text{budget}_C=0.5$ and vary $\text{budget}_A\in[0,1]$.
Fig.~\ref{fig:voting_3groups} shows a monotone trend for the natural and best-response heuristics.
For the softmax heuristic, Spearman's rank correlation is $0.3207$ with $p<0.05$, indicating a weaker but still significant monotone association.

In summary, these simulations suggest that monotonicity of winning probability with respect to a candidate's aggregate advantage
is consistent with a range of simple micro-level voting heuristics.
In particular, the softmax-based heuristics yield smoother curves, 
in the sense that winning probabilities remain strictly between $0$ and $1$ under finite utilities.

\section{Hardness and Counterexamples}
\label{sec:hardness_no_PSNE_examples}

We first note that when the hardmax function is adopted as the monotone WP function, the egoistic election game always admits a PSNE. 
\begin{prop}\label{prop:hardmax_PSNE}
Under the hardmax WP function, every egoistic election game admits a PSNE. 
\end{prop}
\begin{proof}
Firstly, the maximum of a finite set of numbers always exists. By the definition of hardmax, if multiple groups attain the maximum social utility, the winner is the one with the minimum index. That is, candidate $x_{i^*,s_{i^*}}$, where $i^* := \min\argmax_{i\in [m]} u(x_{i,s_i})$ with respect to the profile $\mathbf{s}$, wins with probability~1.
Let 
\[
\mathcal{A}:= \argmax_{i\in[m],\, s\in[n_i]} u(x_{i,s})
\]
be the set of candidates with maximum social utility.
Define
\[
i^*:= \min \{\, i\in[m] : \exists s\in[n_i]\text{ with }(i,s)\in \mathcal{A} \,\},
\]
and choose any $s_{i^*}^*\in[n_{i^*}]$ such that $(i^*,s_{i^*}^*)\in\mathcal{A}$.
Let $I = \{s\in [n_{i^*}]: u(x_{i^*,s})\geq u(x_{j,s_j})\mbox{ for all }j\in [m]\setminus \{i^*\}, s_j\in [n_j]\}$ collect all the candidates in group $i^*$ which bring more social utility than any one does in the other groups. 
Clearly, $I$ is not empty since $s_{i^*}^*\in I$. Next, choose the minimum index  $s_{i^*}^{**}$ in~$I$, which brings the maximum utility for group $i^*$'s members, then we have that any profile $\mathbf{s}$ with group player $i^*$ choosing $s_{i^*}^{**}$ is a PSNE. 
Indeed, any other group player $\mathcal{P}_j$ for $j\neq i^*$ has no incentive to deviate from its strategy because the payoff can never be better off. As for group player $i^*$, by the egoistic property we know that it gets less utility from the other group than that from its own candidates whenever it loses, so choosing $s_{i^*}^{**}$ guarantees his maximum possible reward. 
\end{proof}
Interestingly, a PSNE in the election game where the hardmax WP function is adopted is not necessarily the optimal profile. For example, consider the instance in Table~\ref{tab:AlwaysPSNE_HM}. The social welfare of the PSNE is~$50$, which is only half of the optimum.

\begin{figure}[ht]
    \centering
    \includegraphics[scale=0.50]{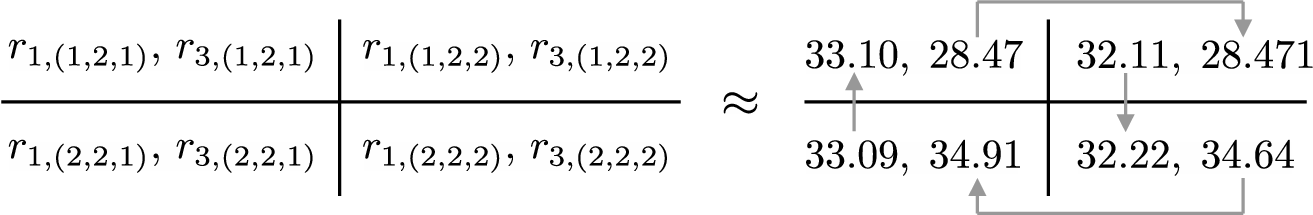}
    \caption{Strategy deviations of the game instance in Table~\ref{tab:NoPSNE_BT}. The arrows indicate the unilateral strategy deviations of~$\mathcal{P}_1$ and $\mathcal{P}_3$.}
    \label{fig:deviation_example_01}
\end{figure}

\begin{figure}[ht]
    \centering
    \includegraphics[scale=0.50]{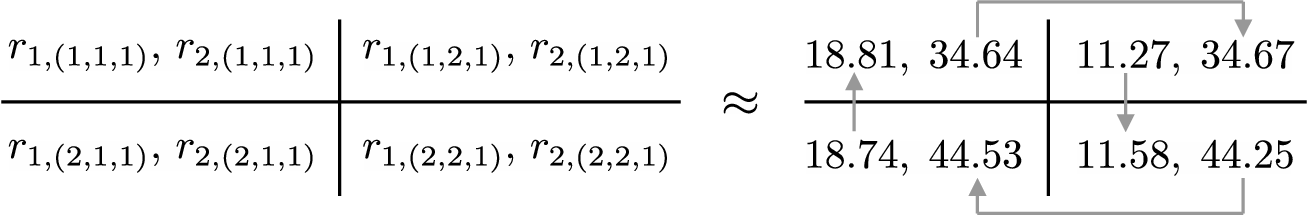}
    \caption{Strategy deviations of the game instance in Table~\ref{tab:NoPSNE_softmax}. The arrows indicate the unilateral strategy deviations of~$\mathcal{P}_1$ and $\mathcal{P}_2$.}
    \label{fig:deviation_example_02}
\end{figure}

Lin et al.~\cite{LLC2021} has shown that the egoistic two-party election game always admits a PSNE if a linear function or the softmax function is adopted as the monotone WP function. 
One might be curious about whether the egoistic property is sufficient for such an election game of \emph{three or more groups} to admit a PSNE. Unfortunately, 
we find game instances as counterexamples, which imply that the egoistic election game of three or more groups does not always admit a PSNE (see Table~\ref{tab:NoPSNE_BT} and 
Table~\ref{tab:NoPSNE_softmax}). 
Indeed, consider the game instance in Table~\ref{tab:NoPSNE_BT}, group player $\mathcal{P}_2$ must play strategy $x_{2,2}$ since
$r_2(i,2,j) > r_2(i,1,j)$ for any $(i,j)\in\{1,2\}^2$. Hence any PSNE must have $\mathcal{P}_2$ playing $x_{2,2}$. Yet, we can observe the strategy deviation cycle shown in Fig.~\ref{fig:deviation_example_01}. Similarly, 
consider the game instance in Table~\ref{tab:NoPSNE_softmax}, group player $\mathcal{P}_3$ has a dominant strategy $x_{3,1}$ since $r_{3,(i,j,1)}>r_{3,(i,j,2)}$ for any $(i,j)\in \{1,2\}^2$. Then, the game instance resembles a two-party election game instance, yet there is also a deviation-cycle, which shows that a PSNE does not exist in this instance (see Fig.~\ref{fig:deviation_example_02}). 
Note that the two-party election game using the softmax function as the monotone WP function has been guaranteed to have a PSNE~\cite{LLC2021}. Unlike the two-party election game case, the increase of winning probability mass due to a unilateral strategy deviation of~$\mathcal{P}_1$ contributes to the decrease of winning probability mass of both~$\mathcal{P}_2$ and $\mathcal{P}_3$. Hence, the analysis in~\cite{LLC2021} does not apply herein.

\begin{table}[ht]
\centering
\footnotesize
\begin{tabular}[c]{ l l l | l l l | l l l }
 $u_1(x_{1,i})$\!\!\!\!\! & $u_2(x_{1,i})$\!\!\!\!\! & $u_3(x_{1,i})$\!\! & \!\!\!
     $u_1(x_{2,i})$\!\!\!\!\! &  $u_2(x_{2,i})$\!\!\!\!\! & $u_3(x_{2,i})$\!\! & \!\!\!
     $u_1(x_{3,i})$\!\!\!\!\! &  $u_2(x_{3,i})$\!\!\!\!\! & $u_3(x_{3,i})$
        \\
	\hline
	50\!\!  &  0\!\!   &  0\!\!  &  15\!\!  &  31\!\!  &  0\!\!  &  10\!\!  &   10\!\!  &  24\!\! \\
        49\!\!  &  29\!\!  &  22\!\! &  16\!\!  &  30\!\!  &  0\!\!  &  10\!\!  &  10\!\!  &  23 \\
	\hline
\end{tabular}
\vspace{7pt}
\begin{tabular}[c]{ l l l | l l l}
	$r_{1,(1,1,1)} = 50$\!\! & $r_{2,(1,1,1)} = 0$ \!\! & $r_{3,(1,1,1)} = 0$\!\! 
     &   $r_{1,(1,1,2)} = 50$\!\! & $r_{2,(1,1,2)} = 0$\!\! & $r_{3,(1,1,2)} = 0$\\
	\hline
	$r_{1,(1,2,1)} = 50$\!\! & $r_{2,(1,2,1)} = 0$\!\! & $r_{3,(1,2,1)} = 0$\!\! 
     &  $r_{1,(1,2,2)} = 50$\!\! & $r_{2,(1,2,2)} = 0$\!\! & $r_{3,(1,2,2)} = 0$\\
\end{tabular}
\begin{tabular}[c]{ l l l | l l l}
	\centering
	$r_{1,(2,1,1)} = 49$\!\! & $r_{2,(2,1,1)} = 29$\!\! & $r_{3,(2,1,1)} = 22$\!\! 
     & $r_{1,(2,1,2)} = 49$\!\! & $r_{2,(2,1,2)} = 29$\!\! &  $r_{3,(2,1,2)} = 22$\\
	\hline
	$r_{1,(2,2,1)} = 49$\!\! & $r_{2,(2,2,1)} =  29$\!\! & $r_{3,(2,2,1)} = 22$\!\! 
     & $r_{1,(2,2,2)} = 49$\!\! & $r_{2,(2,2,2)} = 29$\!\! & $r_{3,(2,2,2)} = 22$\\
\end{tabular}
\vspace{10pt}
\caption{An egoistic election game instance of three groups that admits a PSNE (hardmax WP; $\beta=100, n_i=2$ for $i\in\{1,2,3\}$). For all $i\in[m]$, $\min_{s\in[n_i]} u_i(x_{i,s}) > \max_{j\neq i}\max_{t\in[n_j]} u_i(x_{j,t})$, hence the instance satisfies the egoistic property.}\label{tab:AlwaysPSNE_HM}
\footnotetext{For example, here $u_{x_{1,2}} = 49+29+22 = 100$, $u_{x_{2,2}} = 16+30+0 = 46$, and $u_{x_{3,2}} = 10+10+23 = 43$, so for profile $(2,2,2)$ the winner is $x_{1,2}$ and the rewards of the three group players are $49$, $29$ and $22$ respectively, which result in social welfare $49+29+22 = 100$. By checking all the four profiles we know the optimal profile has social welfare~$100$. It is easy to see that profiles $(1,1,1)$, $(1,1,2)$, $(1,2,1)$ and $(1,2,2)$ are all PSNE.}
\end{table}

\begin{table}[ht]
\centering
\footnotesize
\begin{tabular}[c]{ l l l | l l l | l l l }
$u_1(x_{1,i})$\!\!\!\!\! & $u_2(x_{1,i})$\!\!\!\!\! & $u_3(x_{1,i})$\!\!\! & \!\!\!
         $u_1(x_{2,i})$\!\!\!\!\! & $u_2(x_{2,i})$\!\!\!\!\! & $u_3(x_{2,i})$\!\!\! & \!\!\!
         $u_1(x_{3,i})$\!\!\!\!\! & $u_2(x_{3,i})$\!\!\!\!\! & $u_3(x_{3,i})$
        \\
	\hline
	50  &  9   &  3  &  22  &  42  &  15  &  36  &   1  &  44 \\
        44  &  22  &  30 &  19  &  40  &  32  &  33  &  18  &  42 \\
	\hline
\end{tabular}
\vspace{7pt}
\begin{tabular}[c]{ l l l | l l l}
	$r_{1,(1,1,1)}\approx 34.93$\!\!\! & $r_{2,(1,1,1)}\approx 17.82$\!\!\! & $r_{3,(1,1,1)}\approx 22.23$  
     &  $r_{1,(1,1,2)}\approx 33.79$\!\!\! & $r_{2,(1,1,2)}\approx 23.72$\!\!\! & $r_{3,(1,1,2)}\approx 22.55$\\
	\hline
	$r_{1,(1,2,1)}\approx 33.10$\!\!\! & $r_{2,(1,2,1)}\approx 18.29$\!\!\! & $r_{3,(1,2,1)}\approx 28.47$ 
     &  $r_{1,(1,2,2)}\approx 32.11$\!\!\! & $r_{2,(1,2,2)}\approx 23.87$\!\!\! & $r_{3,(1,2,2)}\approx 28.471$\\
\end{tabular}
\vspace{7pt}
\begin{tabular}[c]{ l l l | l l l}
	$r_{1,(2,1,1)}\approx 34.68$\!\!\! & $r_{2,(2,1,1)}\approx 21.53$\!\!\! & $r_{3,(2,1,1)}\approx 29.80$  
     & $r_{1,(2,1,2)}\approx 33.70$\!\!\! & $r_{2,(2,1,2)}\approx 26.51$\!\!\! & $r_{3,(2,1,2)}\approx 29.74$\\
	\hline
	$r_{1,(2,2,1)}\approx 33.09$\!\!\! & $r_{2,(2,2,1)}\approx 21.76$\!\!\! & $r_{3,(2,2,1)}\approx 34.91$  
     & $r_{1,(2,2,2)}\approx 32.22$\!\!\! & $r_{2,(2,2,2)}\approx 26.52$\!\!\! & $r_{3,(2,2,2)}\approx 34.64$\\
\end{tabular}
\vspace{10pt}
\caption{An egoistic election game instance of three groups that admits no PSNE. The natural function to compute winning probabilities is adopted ($\beta=100, n_i=2$ for $i\in\{1,2,3\}$). For all $i\in[m]$, $\min_{s\in[n_i]} u_i(x_{i,s}) > \max_{j\neq i}\max_{t\in[n_j]} u_i(x_{j,t})$, hence the instance satisfies the egoistic property.}\label{tab:NoPSNE_BT}
\footnotetext{Here, for example, $u_{x_{1,1}} = 50+9+3=62$, $u_{x_{2,1}} = 22+42+15=79$, $u_{x_{3,1}}=36+1+44=81$. Then we have $p_{1,(1,1,1)} = 62/(62+79+81)\approx 0.2793$, $p_{2,(1,1,1)} = 79/(62+79+81)\approx 0.3559$, $p_{3,(1,1,1)}\approx 0.3649$. So $r_{1,1,1} = 50\cdot p_{1,(1,1,1)} + 22\cdot p_{2,(1,1,1)} + 36\cdot p_{3,(1,1,1)}\approx 34.93$. All the eight profiles are not PSNE. For example, consider profile $(1,1,1)$. Group player $\mathcal{P}_2$ has the incentive to change its strategy to~$x_{2,2}$ from~$x_{2,1}$ because $r_{2,(1,2,1)} = 18.29 > 17.82 = r_{2,(1,1,1)}$. This example is clearly egoistic because $u_1(x_{1,2}) = 44 > \max\{u_1(x_{2,1}), u_1(x_{3,1})\} = 36$, $u_2(x_{2,2}) = 40 > \max\{u_2(x_{1,2}), u_2(x_{3,2})\} = 22$ and $u_3(x_{3,2}) = 42 > \max\{u_3(x_{1,2}), u_3(x_{2,2})\} = 32$.}
\end{table}

\begin{table}[ht]
\centering
\footnotesize
\begin{tabular}[c]{ l l l | l l l | l l l }
	{\footnotesize $u_1(x_{1,i})$}\!\!\!\!\! & {\footnotesize $u_2(x_{1,i})$}\!\!\!\!\! & {\footnotesize $u_3(x_{1,i})$}\!\!\! & \!\!\!
        {\footnotesize $u_1(x_{2,i})$}\!\!\!\!\! & {\footnotesize $u_2(x_{2,i})$}\!\!\!\!\! & {\footnotesize $u_3(x_{2,i})$}\!\!\! & \!\!\!
        {\footnotesize $u_1(x_{3,i})$}\!\!\!\!\! & {\footnotesize $u_2(x_{3,i})$}\!\!\!\!\! & {\footnotesize $u_3(x_{3,i})$}
        \\
	\hline
	29  &  4   &  21  &  23  &  59  &  7  &  8  &  32  &  54 \\
        27  &  43  &  3 &  3  &  57  &  38  &  20  &  13  &  53 \\
	\hline
\end{tabular}
\vspace{7pt}
\begin{tabular}[c]{ l l l | l l l}
	\centering
	$r_{1,(1,1,1)}\approx 18.81$\!\!\! & $r_{2,(1,1,1)}\approx 34.64$\!\!\! & $r_{3,(1,1,1)}\approx 28.51$
     & $r_{1,(1,1,2)}\approx 23.49$\!\!\! & $r_{2,(1,1,2)}\approx 27.82$\!\!\! & $r_{3,(1,1,2)}\approx 27.38$\\
	\hline
	$r_{1,(1,2,1)}\approx 11.27$\!\!\! & $r_{2,(1,2,1)}\approx 34.67$\!\!\! & $r_{3,(1,2,1)}\approx 39.70$ 
     & $r_{1,(1,2,2)}\approx 15.57$\!\!\! & $r_{2,(1,2,2)}\approx 28.09$\!\!\! & $r_{3,(1,2,2)}\approx 38.93$\\
\end{tabular}
\vspace{7pt}
\begin{tabular}[c]{ l l l | l l l}
	\centering
	$r_{1,(2,1,1)}\approx 18.74$\!\!\! & $r_{2,(2,1,1)}\approx 44.53$\!\!\! & $r_{3,(2,1,1)}\approx 22.84$ 
     & $r_{1,(2,1,2)}\approx 23.18$\!\!\! & $r_{2,(2,1,2)}\approx 38.35$\!\!\! &  $r_{3,(2,1,2)}\approx 21.61$\\
	\hline
	$r_{1,(2,2,1)}\approx 11.58$\!\!\! & $r_{2,(2,2,1)}\approx 44.25$\!\!\! & $r_{3,(2,2,1)}\approx 33.66$ 
     & $r_{1,(2,2,2)}\approx 15.67$\!\!\! & $r_{2,(2,2,2)}\approx 38.27$\!\!\! & $r_{3,(2,2,2)}\approx 32.77$\\
\end{tabular}
\vspace{10pt}
\caption{An egoistic election game instance of three groups that admits no PSNE. The softmax function to compute winning probabilities is adopted ($\beta=100, n_i=2$ for $i\in\{1,2,3\}$). For all $i\in[m]$, $\min_{s\in[n_i]} u_i(x_{i,s}) > \max_{j\neq i}\max_{t\in[n_j]} u_i(x_{j,t})$, hence the instance satisfies the egoistic property.}\label{tab:NoPSNE_softmax} 
\footnotetext{Here, for example, $u_{x_{1,1}} = 29+4+21=54$, $u_{x_{2,1}} = 23+59+7=89$, $u_{x_{3,1}}=8+32+54=94$. Then we have $p_{1,(1,1,1)} = e^{54/100}/(e^{54/100}+e^{89/100}+e^{94/100})\approx 0.2557$, $p_{2,(1,1,1)} = e^{89/100}/(e^{54/100}+e^{89/100}+e^{94/100})\approx 0.3629$, $p_{3,(1,1,1)}\approx 0.3814$. So $r_{1,1,1} = 29\cdot p_{1,(1,1,1)} + 23\cdot p_{2,(1,1,1)} + 8\cdot p_{3,(1,1,1)}\approx 18.81$. All the eight profiles are not PSNE. For example, consider profile $(1,1,1)$. Group player $\mathcal{P}_2$ has the incentive to change its strategy to~$x_{2,2}$ from~$x_{2,1}$ because $r_{2,(1,2,1)} = 34.67 > 34.64 = r_{2,(1,1,1)}$. This example is egoistic because $u_1(x_{1,2}) = 27 > \max\{u_1(x_{2,1}), u_1(x_{3,2})\} = 23$, $u_2(x_{2,2}) = 57 > \max\{u_2(x_{1,2}), u_2(x_{3,1})\} = 43$ and $u_3(x_{3,2}) = 53 > \max\{u_3(x_{1,1}), u_3(x_{2,2})\} = 38$.}
\end{table}

As an egoistic election game (of three or more groups) does not always admit a PSNE, it is reasonable to investigate the hardness of deciding whether an egoistic election game of $m\geq 2$ group players has a PSNE. As pointed out by~\`{A}lvarez et al.~\cite{AGS2011}, the computational complexity for such a decision problem of equilibrium existence 
can be dependent on 
the degrees of succinctness of the input representation. It is also worth noting that in Sect.~4 of~\cite{GGS05}, Gottlob et al. indicate that to determine if a game in the standard normal form\footnote{A game is in the standard normal form if the payoffs of players can be explicitly represented by a single table or matrix.} has a PSNE can be solved in logarithmic space, and hence in polynomial time although such a representation of a game instance is very space consuming. Instead, we consider the \emph{general form} representation~\cite{AGS2011}, which succinctly represents a game instance in a tuple of players, their action sets and a deterministic algorithm, 
such that the payoffs are not required to be given explicitly. Note that to compute a PSNE of a game under the general form representation is {\sf NP}-hard~\cite{AGS2011}.

We study PSNE existence under a general-form representation of an egoistic election game.
An instance is
\[
\mathcal{I}=(m,\{n_i\}_{i\in[m]},\beta,\{u_j(x_{i,s_i})\}_{i,j\in[m],\,s_i\in[n_i]},\Pi),
\]
where utilities $u_j(x_{i,s_i})$ and $\beta$ are rationals encoded in binary, and $\Pi$ is the WP function 
which is a polynomial-time evaluation procedure 
that, given a profile
$\mathbf{s}=(s_1,s_2,\ldots,s_m)$, outputs winning probabilities $(p_1(\mathbf{s}),p_2(\mathbf{s}),\ldots,p_m(\mathbf{s}))$ and thus the payoffs
\[
r_i(\mathbf{s})=\sum_{j\in[m]} p_j(\mathbf{s})\,u_i(x_{j,s_j}).
\] 
We consider the problem \textsc{EGOISTIC-PSNE-EXIST-GF} formulated as: given $\mathcal{I}$ satisfying the egoistic property,
decide whether a PSNE exists.
Clearly, the input is not given in tabular representation, which requires $\prod_{i\in [m]}n_i = \Omega(2^m)$ entries. Instead, the input is given succinctly and has size~$O(nm^2)$, where $n:=\max_{i\in [m]}n_i$. 
The proof of the {\sf NP}-completeness of determining PSNE existence in an egoistic election game is built upon a reduction from the Satisfiability problem, which resembles the arguments in~\cite{AGS2011} but is revised elaborately to deal with the egoistic property and monotonicity. 

\begin{thm}[{\sf NP}-completeness in general-form representation]\label{thm:NPC}
\textsc{EGOISTIC-PSNE-EXIST-GF} is {\sf NP}-complete, even when $n_i=2$ for all $i\in[m]$, $\beta$ is fixed to $200$, and the instance is monotone. 
\end{thm}
The following Lemma is useful in proving Theorem~\ref{thm:NPC}. 

\begin{lem}[Certified rational interval for $x^{9/10}$]\label{lem:root-interval}
Fix an integer $x\ge 1$ and precision $k\in\mathbb{N}$.
Let $q$ be a nonnegative integer which satisfies 
\[
q^{10}\ \le\ x^9\cdot 10^{10k}\ <\ (q+1)^{10}.
\]
Define $\ell:=q/10^k$ and $u:=(q+1)/10^k$. Then $x^{9/10}\in[\ell,u]$.
\end{lem}
\begin{proof}
The proof is straightforward. Divide the inequalities by $10^{10k}$ to obtain
\[
\left(\frac{q}{10^k}\right)^{10}\ \le\ x^9\ <\ \left(\frac{q+1}{10^k}\right)^{10},
\]
i.e., $\ell^{10}\ \le\ x^9\ <\ u^{10}$.
Since the function $\varphi(t)=t^{10}$ is strictly increasing on $\mathbb{R}_{\ge 0}$, it has a
strictly increasing inverse $\varphi^{-1}(y)=y^{1/10}$ on $\mathbb{R}_{\ge 0}$.
Applying $\varphi^{-1}$ to the inequality yields $\ell\ \leq\ x^{9/10}\ <\ u$. 
\end{proof}

\begin{proof}[Proof of Theorem~\ref{thm:NPC}]
\emph{{\sf NP} membership.}
Given a profile $\mathbf{s}$, the verifier runs $\Pi$ (polynomial time by assumption) to obtain $(p_1(\mathbf{s}),p_2(\mathbf{s}),\ldots,p_m(\mathbf{s}))$
and then computes each $r_i(\mathbf{s})=\sum_j p_j(\mathbf{s})u_i(x_{j,s_j})$ in polynomial time. Note that $\Pi$ does not search over assignments. It only evaluates $F$ on the assignment encoded 
by the profile $\mathbf{s}$, i.e., it computes $F(a(\mathbf{s}))$. 
Hence one can verify whether $\mathbf{s}$ is a PSNE by checking all unilateral deviations, so the problem is in {\sf NP}.

\emph{{\sf NP}-hardness (reduction from \textsc{SAT}).}
Let $F$ be a Boolean formula over variables $v_1,v_2,\ldots,v_n$.
We construct an instance with $m:=n+2$ groups. Intuitively, the first $n$ groups will choose truth values for each variable, and the last two groups will form a small $2\times 2$ game that yields an PSNE if and only if the assignment satisfies the formula. 
For each $i\in[m]$, let group $\mathcal{P}_i$ have two candidates as actions $x_{i,1}$ and $x_{i,2}$.
We interpret, for each $i\in[n]$, action $1$ as setting $v_i=\texttt{True}$ and action $2$ as setting $v_i=\texttt{False}$.
The two groups correspond to variables $v_{n+1}$ and $v_{n+2}$ which do not exist in~$F$ (i.e., the last two groups are not part of the assignment). Set $\beta:=200$. 
Choose any fixed $0<\epsilon<1$. 
As for the utilities, 
for each $i\in[n]$, set $u_i(x_{i,1})=u_i(x_{i,2})=\epsilon$ and $u_j(x_{i,1})=u_j(x_{i,2})=0$ for all $j\neq i$.
For the last two groups, write $m-1=n+1$ and $m=n+2$, and set
\[
\begin{array}{cc}
u_{m-1}(x_{m-1,1}) = 83, & u_{m}(x_{m-1,1}) = 1\\
u_{m-1}(x_{m-1,2}) = 80, & u_{m}(x_{m-1,2}) = 19\\
u_{m-1}(x_{m,1}) = 3, & u_{m}(x_{m,1}) = 24\\
u_{m-1}(x_{m,2}) = 9, & u_{m}(x_{m,2}) = 22
\end{array}
\]
and $u_j(x_{m-1,1})=u_j(x_{m-1,2})=u_j(x_{m,1})=u_j(x_{m,2})=0$ for all $j\leq n$.
This instance is egoistic by construction.
Define the social utilities
$U_{m-1,1}:= u(x_{m-1,1})=84,\; U_{m-1,2}:=u(x_{m-1,2})=99,\; U_{m,1}:= u(x_{m,1}) = 27,\; U_{m,2}:= u(x_{m,2})=31$,
and note that for $i\leq n$ we have $U_{i,1}=U_{i,2}=\epsilon<1$.

\emph{Winning probabilities.} 
Procedure $\Pi$ takes profile $\mathbf{s}=(s_1,s_2,\ldots,s_m)$ and:
(i) interprets $(s_1,s_2,\ldots,s_n)$ as an assignment $a(\mathbf{s})\in\{\texttt{True},\texttt{False}\}^n$ to $(v_1,v_2,\ldots,v_n)$;
(ii) sets
\[
f(\mathbf{s})=\begin{cases}
1 & \text{if }F(a(\mathbf{s}))=\texttt{True},\\
0 & \text{otherwise};
\end{cases}
\qquad
\alpha(\mathbf{s})=1-\frac{f(\mathbf{s})}{10}\in\left\{\frac{9}{10},1\right\};
\]
(iii) outputs winning probabilities
\[
p_i(\mathbf{s})=
\begin{cases}
\dfrac{U_{i,s_i}^{\alpha(\mathbf{s})}}{\sum\limits_{j\in[m]:\,U_{j,s_j}>1} U_{j,s_j}^{\alpha(\mathbf{s})}} & \text{if }U_{i,s_i}>1,\\[2ex]
0 & \text{otherwise}.
\end{cases}
\]
Because evaluating $F$ takes polynomial time in $|F|$ and the remaining arithmetic is constant-time
on fixed-size rationals/integers, $\Pi$ runs in polynomial time and its description size is polynomial in $|F|$.

Since $U_{i,s_i}=\epsilon<1$ for all $i\leq n$, we have $p_i(\mathbf{s})=0$ for $i\leq n$ at every profile $\mathbf{s}$.
Thus only groups $\mathcal{P}_{m-1}$ and $\mathcal{P}_m$ receive positive winning probability, and the strategic interaction
reduces to a $2\times 2$ subgame between these two groups parameterized by $\alpha(\mathbf{s})\in\{\tfrac{9}{10},1\}$.

\emph{Monotonicity.}
For $i\leq n$, switching between candidates keeps total utility at $\epsilon$ and keeps $p_i(\mathbf{s})=0$.
For $i\in\{m-1,m\}$, for any fixed $\alpha>0$ and fixed opponent total utility $U_{\text{opp}}$, the function
$U\mapsto U^\alpha/(U^\alpha+U_{\text{opp}}^\alpha)$ is strictly increasing in $U$ on $\mathbb{R}_{>0}$.
Since $U_{m-1,2}>U_{m-1,1}$ and $U_{m,2}>U_{m,1}$, the constructed instance is monotone.


\paragraph{Case A: $F$ is satisfiable}
Choose any profile $\mathbf{s}$ whose first $n$ components encode a satisfying assignment of~$F$.
Then $f(\mathbf{s})=1$ and $\alpha(\mathbf{s})=9/10$. In this case, denote
\[
A:=84^{9/10},\quad B:=99^{9/10},\quad C:=27^{9/10},\quad D:=31^{9/10},
\]
and for any $\mathbf{s}'$ with $s_{m-1}=i, s_m=j$, denote 
the pair of payoffs $(r_{m-1}(\mathbf{s}'), r_{m}(\mathbf{s}'))$ by $(r_{m-1},r_m)(i,j)$. 
Using the fact that $\beta$ cancels in the probability ratios, the four payoff pairs are exactly:
\[
(r_{m-1},r_m)(1,1)=\left(\frac{83A+3C}{A+C},\ \frac{A+24C}{A+C}\right),\quad
(r_{m-1},r_m)(1,2)=\left(\frac{83A+9D}{A+D},\ \frac{A+22D}{A+D}\right),
\]
\[
(r_{m-1},r_m)(2,1)=\left(\frac{80B+3C}{B+C},\ \frac{19B+24C}{B+C}\right),\quad
(r_{m-1},r_m)(2,2)=\left(\frac{80B+9D}{B+D},\ \frac{19B+22D}{B+D}\right).
\]
We show action $1$ is a strict dominant strategy for both groups $\mathcal{P}_{m-1}$ and~$\mathcal{P}_m$.

\subparagraph{\normalfont (i) Group $\mathcal{P}_{m-1}$ prefers $1$ when $s_m=1$:}
\[
r_{m-1}(1,1)>r_{m-1}(2,1)
\iff (83A+3C)(B+C)>(80B+3C)(A+C)
\iff N_1>0,
\]
where
\[
N_1:=3AB+80AC-77BC.
\]

\subparagraph{\normalfont (ii) Group $\mathcal{P}_{m-1}$ prefers $1$ when $s_m=2$:}
\[
r_{m-1}(1,2)>r_{m-1}(2,2)
\iff (83A+9D)(B+D)>(80B+9D)(A+D)
\iff N_2>0,
\]
where
\[
N_2:=3AB+74AD-71BD.
\]

\subparagraph{\normalfont (iii) Group $\mathcal{P}_m$ prefers $1$ when $s_{m-1}=1$:}
\[
r_{m}(1,1)>r_{m}(1,2)
\iff (A+24C)(A+D)>(A+22D)(A+C)
\iff N_3>0,
\]
where
\[
N_3:=-21AD+23AC+2CD.
\]

\subparagraph{\normalfont (iv) Group $\mathcal{P}_m$ prefers $1$ when $s_{m-1}=2$:}
\[
r_{m}(2,1)>r_{m}(2,2)
\iff (19B+24C)(B+D)>(19B+22D)(B+C)
\iff N_4>0,
\]
where
\[
N_4:=B(5C-3D)+2CD.
\]

Next, we prove certified bounds for $A,B,C,D$ and positivity of~$N_1,N_2,N_3,N_4$. 
We now certify the rational intervals for $A,B,C,D$  
\[
A=84^{9/10},\quad B=99^{9/10},\quad C=27^{9/10},\quad D=31^{9/10}
\]
using Lemma~\ref{lem:root-interval} with $k=2$. 
Specifically, for each $x\in\{84,99,27,31\}$ there exists an integer $q_x$ such that
\[
q_x^{10}\ \leq\ x^9\cdot 10^{20}\ <\ (q_x+1)^{10}.
\]
Applying Lemma~\ref{lem:root-interval} yields $x^{9/10}\in\bigl[q_x/10^2,(q_x+1)/10^2\bigr]$.
For our four values we take
\[
q_{84}=5393,\;\; q_{99}=6252,\;\; q_{27}=1941,\;\; q_{31}=2199,
\]
and thus obtain the certified intervals
\[
A\in[53.93,53.94],\;\; B\in[62.52,62.53],\;\; C\in[19.41,19.42],\;\; D\in[21.99,22.00].
\]
Let $\underline{A}=53.93$, $\overline{A}=53.94$, and similarly
\[
(\underline{B}, \overline{B})=(62.52,62.53),\;\; (\underline{C},\overline{C})=(19.41,19.42),\;\; (\underline{D},\overline{D})=(21.99,22.00).
\]
Since $A,B,C,D>0$, each monomial is increasing in each variable.
Hence, to obtain a \emph{lower bound} on $N_i$ it suffices to evaluate each positive-coefficient
monomial at lower endpoints and each negative-coefficient monomial at upper endpoints. Thus, 
\begin{align*}
& N_1 \geq 3\underline{A}\,\underline{B}+80\underline{A}\,\underline{C}-77\overline{B}\,\overline{C}
> 354,\\ 
& N_2 \geq 3\underline{A}\,\underline{B}+74\underline{A}\,\underline{D}-71\overline{B}\,\overline{D}
> 201,\\
& N_3 \geq -21\overline{A}\,\overline{D}+23\underline{A}\,\underline{C}+2\underline{C}\,\underline{D}
> 9,\\
& N_4 \geq 5\underline{B}\,\underline{C}-3\overline{B}\,\overline{D}+2\underline{C}\,\underline{D}
> 2794.
\end{align*}
Therefore $N_1,N_2,N_3,N_4>0$, so all four best-response inequalities are strict.
It follows that action~$1$ is a strict dominant strategy for both groups $\mathcal{P}_{m-1}$ and~$\mathcal{P}_m$,
and hence $(s_{m-1},s_m)=(1,1)$ is a PSNE of the $2\times 2$ subgame when $\alpha=9/10$.

\paragraph{Case B: $F$ is unsatisfiable}
If $F$ is unsatisfiable, then $f(\mathbf{s})=0$ for all profiles $\mathbf{s}$, hence
$\alpha(\mathbf{s})=1$ always. In this case the weights are linear in total utilities.
Since $u(x_{i,s_i})=\epsilon<1$ for all $i\leq n$, we have $p_i(\mathbf{s})=0$ for $i\leq n$ at every profile.
Thus only groups $\mathcal{P}_{m-1}$ and $\mathcal{P}_m$ have positive winning probabilities, and the induced interaction
is a $2\times 2$ subgame between them.

For $(s_{m-1},s_m)\in\{1,2\}^2$, let $U_A:=u(x_{m-1,s_{m-1}})$ and $U_B:=u(x_{m,s_m})$.
When $\alpha=1$, the winning probabilities are
\[
p_{m-1}=\frac{U_A}{U_A+U_B},\qquad p_m=\frac{U_B}{U_A+U_B}.
\]
Therefore the payoffs are
\[
r_{m-1}=\frac{U_A\cdot u_{m-1}(x_{m-1,s_{m-1}})+U_B\cdot u_{m-1}(x_{m,s_m})}{U_A+U_B},\quad
r_{m}=\frac{U_A\cdot u_{m}(x_{m-1,s_{m-1}})+U_B\cdot u_{m}(x_{m,s_m})}{U_A+U_B}.
\]

Substituting $(U_{m-1,1},U_{m-1,2},U_{m,1},U_{m,2})=(84,99,27,31)$ and the individual utilities yields:
\[
(r_{m-1},r_m)(1,1)=\left(\frac{84\cdot 83+27\cdot 3}{84+27},\frac{84\cdot 1+27\cdot 24}{84+27}\right)
=\left(\frac{2351}{37},\frac{244}{37}\right),
\]
\[
(r_{m-1},r_m)(1,2)=\left(\frac{84\cdot 83+31\cdot 9}{84+31},\frac{84\cdot 1+31\cdot 22}{84+31}\right)
=\left(\frac{7251}{115},\frac{766}{115}\right),
\]
\[
(r_{m-1},r_m)(2,1)=\left(\frac{99\cdot 80+27\cdot 3}{99+27},\frac{99\cdot 19+27\cdot 24}{99+27}\right)
=\left(\frac{127}{2},\frac{281}{14}\right),
\]
\[
(r_{m-1},r_m)(2,2)=\left(\frac{99\cdot 80+31\cdot 9}{99+31},\frac{99\cdot 19+31\cdot 22}{99+31}\right)
=\left(\frac{8199}{130},\frac{2563}{130}\right).
\]

Thus best responses strictly cycle due to the following observations:
\[
\text{Since }\frac{766}{115} > \frac{244}{37}, \text{ we know that }
\text{if }s_{m-1}=1,\ \text{$\mathcal{P}_m$ strictly prefers } s_{m} = 2;
\]
\[
\text{since }\frac{281}{14} > \frac{2563}{130}, \text{ we know that }
\text{if }s_{m-1}=2,\ \text{$\mathcal{P}_m$ strictly prefers } s_m = 1;
\]
\[
\text{since }\frac{2351}{37}-\frac{127}{2}, \text{ we know that }
\text{if }s_m=1,\ \text{$\mathcal{P}_{m-1}$ strictly prefers } s_{m-1} = 1;
\]
\[
\text{since }\frac{8199}{130}>\frac{7251}{115}, \text{ we know that }
\text{if }s_m=2,\ \text{$\mathcal{P}_{m-1}$ strictly prefers }s_{m-1} = 2.
\]
Hence the $2\times 2$ subgame has no PSNE when $\alpha=1$.

Finally, because $p_i(\mathbf{s})=0$ for all $i\leq n$ at every profile, groups $\mathcal{P}_1,\mathcal{P}_2,\ldots,\mathcal{P}_n$ have payoff
$r_i(\mathbf{s})=0$ for all $\mathbf{s}$ and any unilateral deviation by them does not affect
the payoffs of groups $\mathcal{P}_{m-1}$ and~$\mathcal{P}_m$. Therefore no profile of the full game can be a PSNE,
and the constructed instance admits no PSNE.

We have constructed $\mathcal{I}(F)$ in polynomial time such that $\mathcal{I}(F)$ has a PSNE
iff $F$ is satisfiable. Therefore, \textsc{EGOISTIC-PSNE-EXIST-GF} is {\sf NP}-complete.
\end{proof}

\paragraph{Remark} Our {\sf NP}-completeness result (i.e., Theorem~\ref{thm:NPC}) applies to the general-form 
setting where the WP function is given as part of the input. If the WP function is fixed in advance to a 
particular rule (e.g., hardmax/softmax), the computational complexity of PSNE existence may change, 
and {\sf NP}-completeness does not follow automatically from our result. Indeed, when the WP function is 
hardmax, we have shown that the problem can be decided in polynomial time. For the softmax rule defined 
in Sect.~\ref{sec:preliminaries}, whether PSNE existence is {\sf NP}-complete remains open 
and is not implied by Theorem~\ref{thm:NPC}.


\section{Fixed-Parameter Tractable Results}
\label{sec:fpt-algo}

\subsection{Parameterized Problems and Algorithms}

In this subsection, we provide formal definitions of parameterized problems and fixed-parameter tractability to make our discussions self-contained. 

\begin{defn}[Parameterized Problem~\cite{DF13,FG06,Nie06}]
\label{defn:parameterized_problems}
A parameterized problem is a language $\mathcal{L}\subseteq \Sigma^* \times \mathbb{N}^t$ for some fixed constant $t\geq 1$, where $\Sigma$ is a finite alphabet. The second component is called the parameter(s) of the problem.
\end{defn}

\begin{defn}[Fixed-Parameter Tractability~\cite{DF13,FG06,Nie06}]
\label{defn:fpt}
A parameterized problem $\mathcal{L}$ is fixed-parameter tractable ({\sf FPT}) if determining whether $(x,\mathbf{k})\in \mathcal{L}$ can be done in $f(\mathbf{k})\,n^{O(1)}$ time, where $\mathbf{k}\in\mathbb{N}^t$, $n = |x|$ denotes the encoding length of~$x$, and $f$ is a computable function that depends only on~$\mathbf{k}$.
\end{defn}

\subsection{Sufficient Conditions and the Algorithm}
\label{subsec:fpt_condidtions}

Theorem~\ref{thm:dominatingNE} provides two sufficient conditions for the egoistic election game to admit a PSNE when a cross-monotone WP function is adopted. Roughly speaking, if every group player has a dominant strategy by nominating their first candidate, which weakly surpasses all the other ones in their own candidate sets, then clearly the profile is a PSNE. Moreover, if $m-1$ of the group players have such dominant strategies, the remaining group player can simply take a best response with respect to the profile except itself. 

\begin{thm}\label{thm:dominatingNE}
Consider an egoistic election game instance with a \emph{cross-monotone} WP function. 
\begin{enumerate}[label=(\alph*)]
\item If for all $i\in [m]$, strategy $x_{i,1}$ weakly surpasses (surpasses, resp.) every $x_{i,j}$ for $j\in [n_i]\setminus\{1\}$ of group player~$\mathcal{P}_i$, then $(x_{1,1}, x_{2,1}, \ldots, x_{m,1})$ is a PSNE (the unique PSNE, resp.) of the egoistic election game. 
\item Suppose that there exists $\mathcal{I}\subset [m]$, $|\mathcal{I}| = m-1$, such that for all $i\in \mathcal{I}$,  
$x_{i,1}$ weakly surpasses every $x_{i,j}$ for $j\in [n_i]\setminus\{1\}$ of group player $\mathcal{P}_i$. 
Let $i'$ be the unique element of $[m]\setminus \mathcal{I}$.
Then $((x_{i,1})_{i\in\mathcal{I}}, x_{i',s_{i'}^{\#}})$ is a PSNE where $s_{i'}^{\#} \in \argmax_{\ell\in [n_{i'}]} r_{i'}((x_{i,1})_{i\in\mathcal{I}}, x_{i',\ell})$. 
\end{enumerate}
\end{thm}
\begin{proof}
For the first case, let us consider an arbitrary $i\in [m]$ and an arbitrary $t\in [n_i]\setminus\{1\}$. Denote by $\mathbf{s}$ the profile $(x_{1,1}, x_{2,1}, \ldots, x_{m,1})$. 
By the egoistic property, we have $u_i(x_{j,1}) < u_i(x_{i,t})$ for every $j\neq i$.
Moreover, since $x_{i,1}$ weakly surpasses $x_{i,t}$, we have $u(x_{i,1}) \ge u(x_{i,t})$
and thus $p_{i,\mathbf{s}} \ge p_{i,(t,\mathbf{s}_{-i})}$ by monotonicity.
Because the WP function is \emph{cross-monotone}, this further implies
$p_{j,\mathbf{s}} \le p_{j,(t,\mathbf{s}_{-i})}$ for all $j\neq i$,
Hence, 
\begin{eqnarray*}
& &r_{i,\mathbf{s}} - r_{i,(t, \mathbf{s}_{-i})}\\ 
&=& 
\sum_{j\in [m]} p_{j,\mathbf{s}} u_i(x_{j,1}) - \Big(\sum_{j\in [m]\setminus\{i\}} p_{j,(t, \mathbf{s}_{-i})} u_i(x_{j,1}) + p_{i,(t, \mathbf{s}_{-i})} u_i(x_{i,t}) \Big)\\
&=& \sum_{j\in [m]\setminus\{i\}} u_i(x_{j,1})(p_{j,\mathbf{s}}-p_{j,(t,\mathbf{s}_{-i})}) + (p_{i,\mathbf{s}} u_i(x_{i,1}) - p_{i,(t,\mathbf{s}_{-i})} u_i(x_{i,t}))\\
&\geq& \sum_{j\in [m]\setminus\{i\}} u_i(x_{j,1})(p_{j,\mathbf{s}}-p_{j,(t,\mathbf{s}_{-i})}) +  u_i(x_{i,t}) (p_{i,\mathbf{s}} - p_{i,(t,\mathbf{s}_{-i})})\\
&>& u_i(x_{i,t})\Big( \sum_{j\in [m]\setminus\{i\}} p_{j,\mathbf{s}} - p_{j,(t,\mathbf{s}_{-i})} + (p_{i,\mathbf{s}} - p_{i,(t,\mathbf{s}_{-i})})\Big)\\
&=& u_i(x_{i,t})\Big(\sum_{j\in [m]} p_{j,\mathbf{s}} - \sum_{j\in [m]} p_{j,(t,\mathbf{s}_{-i})} \Big)\\
&=& 0, 
\end{eqnarray*}
where the inequality follows from the assumption that strategy~$x_{i,1}$ weakly surpasses~$x_{i,t}$, the second inequality follows from the egoistic property and assumption that the WP function is cross-monotone, and the last equality holds since $\sum_{j\in [m]} p_{j,\mathbf{s}} = \sum_{j\in [m]} p_{j,(t,\mathbf{s}_{-i})} = 1$ by the law of total probability. Thus, group player~$\mathcal{P}_i$ has no incentive to deviate from its current strategy. Note that the uniqueness of the PSNE comes when the inequality becomes ``greater-than''. This happens as strategy~$x_{i,1}$ surpasses~$x_{i,t}$. 

For the second case, by the same arguments for the first case, we know that for $i\in \mathcal{I}$, group player~$\mathcal{P}_i$ has no incentive to deviate from strategy~1. Let $i' = [m]\setminus\mathcal{I}$ be the only one group player not in~$\mathcal{I}$. As $s_{i'}^{\#}$ is the best response when the other strategies of groups in~$\mathcal{I}$ are fixed, group player~$\mathcal{P}_{i'}$ has no incentive to deviate from its current strategy. Therefore, the theorem is proved. 
\end{proof}
For example, consider $m=2$ (i.e., only two groups exist in the organization). Namely, if strategy $x_{1,1}$ of group $\mathcal{P}_1$ (weakly) surpasses each $x_{1,t}$ for $2\leq t\leq n_1$ and strategy $x_{2,1}$ of group $\mathcal{P}_2$ (weakly) surpasses each $x_{2,t'}$ for $2\leq t'\leq n_2$, then $(x_{1,1},x_{2,1})$ is a (weakly) dominant-strategy solution.


We next establish a simple elimination rule that will be used in this and next sections.  
Under cross-monotone WP functions (Definition~\ref{def:crossmono}), any profile that contains a surpassed strategy is not a PSNE.
 
\begin{lem}\label{lem:dominated_Not_PSNE}
Consider an egoistic election game instance with a cross-monotone WP function.
If a group player $\mathcal{P}_i$ has a strategy $s_i$ that is surpassed by another strategy~$s'_i\in [n_i]\setminus\{s_i\}$, then $(s_i,(\tilde{s}_j)_{j\in [m]\setminus\{i\}})$ is not a PSNE for any 
$(\tilde{s}_j)_{j\in [m]\setminus\{i\}}$.
\end{lem}
\begin{proof}
Let $\mathbf{s}_{-i} = (\tilde{s}_j)_{j\in [m]\setminus\{i\}}$ be any profile except group player $\mathcal{P}_i$'s strategy. 
If $s_i$ is surpassed by $s'_i$, then we have $u(x_{i,s'_i})\geq u(x_{i,s_i})$ and $u_i(x_{i,s'_i})\geq u_i(x_{i,s_i})$, and either
\begin{itemize}
\item [(a)] $p_{i, (s_i',\mathbf{s}_{-i})} > p_{i, (s_i, \mathbf{s}_{-i})}$, or
\item [(b)] $p_{i, (s_i, \mathbf{s}_{-i})} > 0$ and $u_i(x_{i,s'_i})>u_i(x_{i,s_i})$. 
\end{itemize}
We prove case (a) as follows. 
By the cross-monotonicity, if $p_{i,(s_i',\mathbf{s}_{-i})} > p_{i,(s_i,\mathbf{s}_{-i})}$, then
$p_{j,(s_i',\mathbf{s}_{-i})} \leq p_{j,(s_i,\mathbf{s}_{-i})}$ for all $j\neq i$. Thus,
\begin{eqnarray*}
& & r_{i}(s_i,\mathbf{s}_{-i}) - r_{i}(s'_i,\mathbf{s}_{-i})\\ 
&=& \Big(p_{i,(s_i,\mathbf{s}_{-i})}u_i(x_{i,s_i}) +\!\!\!\!\! \sum_{j\in [m]\setminus\{i\}} p_{j,(s_i,\mathbf{s}_{-i})}u_i(x_{j,\tilde{s}_j})\Big)\\  
& & - \Big(p_{i,(s'_i,\mathbf{s}_{-i})}u_i(x_{i,s'_i})+\!\!\!\!\!\sum_{j\in [m]\setminus\{i\}} p_{j,(s'_i,\mathbf{s}_{-i})}u_i(x_{j,\tilde{s}_j})\Big)\\
&=& \left(\sum_{j\in [m]\setminus\{i\}} (p_{j,(s_i,\mathbf{s}_{-i})} - p_{j,(s'_i,\mathbf{s}_{-i})}) u_i(x_{j,\tilde{s}_j})\right) 
 + (p_{i,(s_i,\mathbf{s}_{-i})}u_i(x_{i,s_i}) - p_{i,(s'_i,\mathbf{s}_{-i})}u_i(x_{i,s'_i}))\\
&<& u_i(x_{i,s_i})\left(\sum_{j\in [m]\setminus\{i\}} (p_{j,(s_i,\mathbf{s}_{-i})} - p_{j,(s'_i,\mathbf{s}_{-i})})\right) 
+ (p_{i,(s_i,\mathbf{s}_{-i})}u_i(x_{i,s_i}) - p_{i,(s'_i,\mathbf{s}_{-i})}u_i(x_{i,s'_i}))\\
&=& u_i(x_{i,s_i})(p_{i,(s'_i,\mathbf{s}_{-i})} - p_{i,(s_i,\mathbf{s}_{-i})}) + (p_{i,(s_i,\mathbf{s}_{-i})}u_i(x_{i,s_i})
- p_{i,(s'_i,\mathbf{s}_{-i})}u_i(x_{i,s'_i}))\\
&<& (p_{i,(s'_i,\mathbf{s}_{-i})} - p_{i,(s_i,\mathbf{s}_{-i})})(u_i(x_{i,s_i})-u_i(x_{i,s_i}))\\
&=& 0,
\end{eqnarray*}
where the first inequality follows from the egoistic property and the second one follows from the assumption of case (a) and that all coefficients $p_{j,(s_i,\mathbf{s}_{-i})}-p_{j,(s_i',\mathbf{s}_{-i})}$ are nonnegative. 
Thus, $(s_i,\mathbf{s}_{-i})$ is not a PSNE. 
Case (b) can be similarly proved. In particular, monotonicity of the WP function and
$u(x_{i,s_i'})\ge u(x_{i,s_i})$ imply $p_{i',(s_{i'}, \mathbf{s}_{-i})} \geq p_{i, (s_{i}, \mathbf{s}_{-i})} > 0$, and the strict inequality
$u_i(x_{i,\mathbf{s}_{i'}}) > u_i(x_{i,\mathbf{s}_i})$ guarantees $r_i(s_i',\mathbf{s}_{-i}) > r_i(s_i,\mathbf{s}_{-i})$.
\end{proof}

Inspired by Theorem~\ref{thm:dominatingNE} and Lemma~\ref{lem:dominated_Not_PSNE}, we are able to devise an efficient algorithm to find out a PSNE of the egoistic election game with a cross-monotone WP function whenever it exists, with respect to two parameters: \emph{number of irresolute groups} and \emph{nominating depth}, which are introduced as follows. As we have assumed, candidates in each group are sorted according to the utility for its members. That is, $u_i(x_{i,1})\geq u_i(x_{i,2})\geq \ldots \geq u_i(x_{i,n_i})$ for each $i\in [m]$. Let $d_i$ be the index of the candidate that surpasses all candidates $x_{i,d_i+1},\ldots,x_{i,n_i}$ or is set to~$n_i$ if $x_{i,n_i}$ is not surpassed by any candidates of~$\mathcal{P}_i$. Formally, let $\mbox{maxProb}_i := \argmax_{s\in [n_i]}u(x_{i,s})$, then $d_i = \max\{ \arg\max_{s\in \mbox{\scriptsize maxProb}_i} u_i(x_{i,s})\}$. 
From Theorem~\ref{thm:dominatingNE} we know that $x_{i,d_i}$ is a dominant strategy for the ``sub-game instance'' in which $\mathcal{P}_i$'s strategy set is reduced to~$\{x_{i,s}\}_{s\in\{d_i,d_i+1,\ldots,n_i\}}$. Thus, $\mathcal{P}_i$ suffices to consider candidates in $\{x_{i,s}\}_{s\in [d_i]}$ as its strategies. 
We then collect the set $\mathcal{D} = \{i\mid i\in [m], d_i = 1\}$. 
For each $i\in\mathcal{D}$, group player $\mathcal{P}_i$ must choose the first candidate $x_{i,1}$. 
Thus, we can reduce the game instance to $(\tilde{X}_i)_{i\in [m]\setminus\mathcal{D}}$, where $\tilde{X}_i = \{x_{i,s}\}_{s\in [d_i]}$. We call  $d:=\max_{i\in [m]} d_i$ the \emph{nominating depth} of the election game. We call a group $\mathcal{P}_i$ \emph{irresolute} if $d_i>1$ and denote by~$k$ the number of irresolute groups. 
Then, we propose Algorithm {\sf FPT-ELECTION-PSNE} to compute a PSNE of the egoistic election game. The complexity of the algorithm is $O(nm^2 + kd^{k+1}m)$. Thus, to compute a PSNE for such a game is \emph{fixed-parameter tractable} with respect to the parameters~$d$ and~$k$. Theorem~\ref{thm:PSNE_compute_tractable} concludes the results. 

\paragraph{Remark on Positioning and novelty of Algorithm {\sf FPT-ELECTION-PSNE}}
Algorithm {\sf FPT-ELECTION-PSNE} is inspired by elimination of dominated strategies in normal-form games (see~\cite{OR1994CourseGT,CS2005IteratedDominance,GKZ1993EliminatingDominated}). 
However, our pruning criterion is tailored to succinct election games specified via a cross-monotone WP function:
we remove within-group candidates that are \emph{surpassed}, which certifies that such candidates cannot appear in any PSNE (Lemma~\ref{lem:dominated_Not_PSNE}).
This differs from standard dominance, which is defined profile-independently over opponents' actions, whereas surpassing leverages the cross-monotonicity (and hence monotonicity) of the WP function and the election structure.
After pruning, we enumerate the remaining profiles over the $k$ unresolved groups (at most $d^k$ profiles) and verify the PSNE conditions directly, yielding an {\sf FPT} algorithm parameterized by $(k,d)$.

\begin{algorithm}\label{alg:fpt_algo}
\caption{\!$\textbf{\sf FPT-ELECTION-PSNE}$}
\begin{algorithmic}[0]
\REQUIRE an election game instance $\mathcal{G} = (X_1,X_2,\ldots,X_m,f_{\mathcal{G}})$. 
\end{algorithmic}
\begin{algorithmic}[1]
\STATE For each $i$, compute $\mbox{maxProb}_i = \argmax_{s\in [n_i]}u(x_{i,s})$. 
\STATE For each $i$, compute $d_i = \max\{ \arg\max_{s\in \mbox{\scriptsize maxProb}_i} u_i(x_{i,s})\}$.  
\STATE Collect $\mathcal{D} = \{i\mid i\in [m], d_i = 1\}$ and assign $x_{i,1}$ to group player $\mathcal{P}_i$ for $i\in\mathcal{D}$, 
\STATE Reduce the game instance to $(\tilde{X}_i)_{i\in [m]\setminus\mathcal{D}}$, where $\tilde{X}_i = \{x_{i,s}\}_{s\in [d_i]}$, $i\in [m]\setminus\mathcal{D}$. 
\FOR{each entry $\mathbf{s}\in \prod_{i\in \mathcal{D}}\{x_{i,1}\}\times\prod_{j\in [m]\setminus\mathcal{D}} \tilde{X}_j$}
\STATE Compute the payoff $r_i(\mathbf{s})$ for each $i$.
\STATE \COMMENT{Then, check if any unilateral deviation is possible}
\IF{$\mathbf{s}$ corresponds to a PSNE}
\STATE \COMMENT{\!i.e., for each $j\!\in\! [m]\!\setminus\!\mathcal{D}$, check if $r_j(\mathbf{s})\!\geq\! r_j(s'_j,\mathbf{s}_{-j})$ for all $s'_j\!\in\! \tilde{X}_j$}
\STATE return $\mathbf{s}$
\ENDIF
\ENDFOR
\STATE Output ``NO"
\end{algorithmic}
\end{algorithm}

\begin{thm}\label{thm:PSNE_compute_tractable}
Given an election game instance $\mathcal{G}$ of $m\geq 2$ groups each of which has at most $n$ candidates. Assume the WP function is cross-monotone. Suppose that $\mathcal{G}$ has at most~$k$ irresolute groups and the nominating depth of~$\mathcal{G}$ is bounded by~$d$, then to compute a PSNE of~$\mathcal{G}$ takes $O(nm^2 + kd^{k+1}m)$ time if it exists.     
\end{thm}
\begin{proof}
It costs $O(nm^2)$ time to compute $d_i$ for all~$i$ and $d = \max_{i\in [m]} d_i$, and then the set $\mathcal{D}$ can be obtained. Since playing strategy~1 (i.e., nominating the first candidate) for groups $i\in\mathcal{D}$ is the dominant strategy, we only need to consider group players in~$[m]\setminus\mathcal{D}$. Since for each $i\in [m]\setminus\mathcal{D}$, each strategy in~$\{x_{i,d_i+1},\ldots,x_{i,n_i}\}$ is surpassed by $x_{i,d_i}$ (considering the subgame with respect to~$(x_{i,d_i+1},x_{i,d_i+2},\ldots,x_{i,n_i})_{i\in [m]}$), by Lemma~\ref{lem:dominated_Not_PSNE} we know that we only need to consider strategies $\{x_{i,1},x_{i,2},\ldots,x_{i,d_i}\}$ for such group player~$\mathcal{P}_i$. 
Thus, the number of profiles we enumerate is
\[
\prod_{i\in [m]\setminus\mathcal{D}} d_i = O(d^k),
\]
where $k:= |[m]\setminus\mathcal{D}|$ and $d= \max_{i\in[m]\setminus\mathcal{D}} d_i$.
For each enumerated profile $\mathbf{s}$ of the $k$ players in $[m]\setminus\mathcal{D}$ with all other players fixed as specified in the theorem, we first compute the winning probabilities and the payoffs of all players. 
This takes $O(m)$ time under our WP representation. 

We then verify whether $\mathbf{s}$ is a PSNE by checking unilateral deviations of each player $i\in[m]\setminus\mathcal{D}$:
for every alternative strategy $t\in[d_i]\setminus\{s_i\}$, we evaluate the payoff of $i$ under the deviating profile $(t,\mathbf{s}_{-i})$
and compare it with its payoff under $\mathbf{s}$.
The number of such deviation checks is
\[
\sum_{i\in[m]\setminus\mathcal{D}} (d_i-1) \;\le\; k(d-1).
\]
Each deviation check requires recomputing the deviator's payoff and the relevant winning probabilities, which costs $O(m)$ time.
Therefore, checking whether a fixed $\mathbf{s}$ is a PSNE takes $O(m\cdot k(d-1))$ time, and the total running time is $O(nm^2 + d^k \cdot m\cdot k(d-1)) = O(nm^2 + kd^{k+1}m)$. 
\end{proof}

\paragraph{Remark} As candidates in each group are sorted by the utility provided for their own group members, 
the nominating depth~$d$ can be viewed as the maximum number of top-ranked candidates that it suffices to consider per group. 
On the other hand, a group is irresolute if it does not have a candidate as a dominant strategy, and hence a ``searching'' 
for the state involving irresolute groups is required. 
When the number~$k$ of irresolute groups is small, the exponential factor in the running time remains controlled. 
Moreover, in many applications each group has only a small number of plausible nominees, suggesting that both parameters
$(k,d)$ are typically small. Consequently, the proposed fixed-parameter algorithm is likely to be efficient in such settings.   

\subsection{Shrinking the Nomination Depth for Each Group Player}
\label{subsec:strategy_refined}

\begin{figure}[ht]
    \centering
    \includegraphics[scale=0.40]{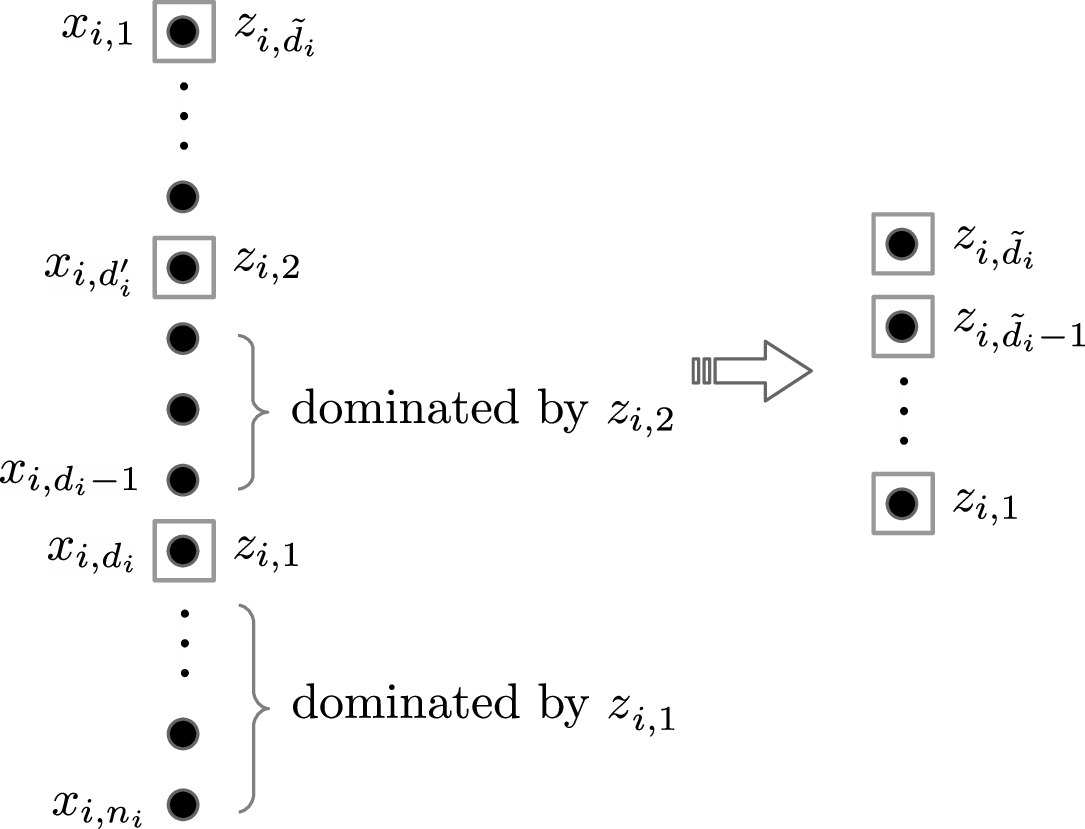}
    \caption{Reducing the strategy set for $\mathcal{P}_i$.}
    \label{fig:strategy_reduce}
\end{figure}

In fact, we can skip some candidates in~$[d_i]$ for each $\mathcal{P}_i$. First, we denote by $L_i^{0}$ the set $[n_i]$ for group player $\mathcal{P}_i$. 
Recall that $x_{i,d_i}$ is a candidate which surpasses the candidates in $\{x_{i, d_i+1},x_{i, d_i+2},\ldots,x_{i, n_i}\}$ if $d_i\neq n_i$ otherwise $x_{i, d_i} = x_{i,n_i}$. Let us denote $x_{i,d_i}$ by~$z_{i,1}$ to facilitate our discussion. We can apply the same approach to identify the candidate $z_{i,2} = x_{i,d'_i}$ if there exists $d'\in [1, d_i-1)$ such that $x_{i,d'_i}$ surpasses candidates in~$\{x_{i,d_i'+1},x_{i,d_i'+2},\ldots, x_{i,d_i-1}\}$ and set $z_{i,2} = x_{i,d_i-1}$ otherwise (see Figure~\ref{fig:strategy_reduce}). 
By repeatedly computing $z_{i,1}, z_{i,2},\ldots,z_{i,\tilde{d}_i}$, in which $\tilde{d}_i$ is the number of proceeded repetitions and $z_{i,\tilde{d}_i} = x_{i,1}$. Clearly, $\tilde{d}_i$ is upper bounded by~$d_i$. As it suffices to consider candidates that are not dominated by any other one for each group to seek a PSNE, we can consider $\{z_{i,1},z_{i,2},\ldots,z_{i, \tilde{d}_i}\}$ as the strategy set of $\mathcal{P}_i$. 

The nominating depth $d = \max_{i\in [m]}\{d_i\}$ can be substituted by $\tilde{d} := \max_{i\in [m]} \tilde{d}_i$. Clearly, $\tilde{d}\leq d$. Thus, using the refined strategy set $\{\hat{s}_1,\hat{s}_2,\ldots,\hat{s}_{\tilde{d}_i}\}$ for each group player, $\mathcal{P}_i$ leads to a possibly more efficient {\sf FPT} algorithm (with a possibly smaller exponential base~$\tilde{d}$), though the time complexity in the worst case still coincides.

\section{Price of Anarchy}
\label{sec:PoA}

Below, Proposition~\ref{pro:cases} relates a PSNE to an optimal profile on social welfare. Note that a PSNE may be a profile that is suboptimal in the social welfare in the election game.


\begin{prop}\label{pro:cases}
Let $\mathbf{s} = (s_i)_{i\in [m]}$ be a PSNE and $\mathbf{s}^* = (s_i^*)_{i\in [m]}$ be the optimal profile which has the highest social welfare among all possible profiles. Then, $\sum_{i\in [m]}u(s_i)\geq \max_{i\in [m]} u(s_i^*)$.
\end{prop}
\begin{proof}
We assume that $\mathbf{s}\neq \mathbf{s}^*$ since otherwise the proposition trivially holds. By Lemma~\ref{lem:dominated_Not_PSNE}, we know that for each $i\in [m]$, strategy $s_i$ is not surpassed by~$s_i^*$ since $\mathbf{s}$ is a PSNE. Therefore, it suffices to consider the following cases.
\begin{enumerate}[label=(\alph*)]
    \item If $s_i\leq s_i^*$ for all $i\in [m]$, then for each $j\in [m]$, 
    $\sum_{i\in [m]} u(s_i)\geq \sum_{i\in [m]}u_i(s_i)\geq \sum_{i\in [m]} u_i(s_j^*) = u(s_j^*)$, where the second inequality follows from the egoistic property. Hence, we have $\sum_{i\in [m]}u(s_i)\geq \max_{i\in [m]} u(s_i^*)$.  
    \item If $u(s_i^*)\leq u(s_i)$ for all $i\in [m]$, then obviously we have 
	$\sum_{i\in [m]}u(s_i)\geq \sum_{i\in [m]}u(s_i^*)\geq \max_{i\in [m]}u(s_i^*)$. 
    \item Suppose that there exists a subset $W\subset [m]$ such that $s_i\leq s_i^*$ for all $i\in W$ and $u(s_j^*)\leq u(s_j)$ for each $j\in \overline{W}:=[m]\setminus W$. For $j\in W$, we have $\sum_{i\in [m]} u(s_i)\geq \sum_{i\in [m]} u_i(s_i)\geq \sum_{i\in [m]}u_i(s_j^*) = u(s_j^*)$. For $j\in\overline{W}$, we have $\sum_{i\in [m]} u(s_i) = \sum_{i\in [m]\setminus\{j\}} u(s_i) + u(s_j)\geq u(s_j^*)$. Hence, $\sum_{i\in [m]} u(s_i)\geq \max_{i\in [m]}u(s_i^*)$.
\end{enumerate}
\end{proof}

By carefully lower bounding the social welfare of a profile, we can derive the following bounds for the egoistic election game with any cross-monotone and order-preserving WP function. 
For any strategy profile $\mathbf{s}$, we have
\vspace{-0pt}
\begin{align}
& SW(\mathbf{s}) = \sum_{i\in [m]} p_{i,\mathbf{s}}\cdot u(s_i)\leq \max_{i\in [m]} u(s_i) \label{eq:UB_SW_softmax}\\
& SW(\mathbf{s}) = \sum_{i\in [m]} p_{i,\mathbf{s}}\cdot u(s_i) \geq \frac{1}{m}\cdot \sum_{i\in [m]} u(s_i) \label{eq:LB_SW_softmax}
\end{align} 
where inequality~\ref{eq:UB_SW_softmax} holds for any WP function, and inequality~\ref{eq:LB_SW_softmax} holds for any order-preserving WP function. 
Inequality~(\ref{eq:LB_SW_softmax}) can be justified as follows. 
W.l.o.g., let us assume that $u(s_1)\geq u(s_2)\geq \ldots \geq u(s_m)$ (by relabeling after they are sorted). As the winning probability function is order-preserving, it is clear that $p_{1,\mathbf{s}}\geq p_{2,\mathbf{s}}\geq\ldots \geq p_{m,\mathbf{s}}$. Let $k\in [m]$ be the index such that $p_{k,\mathbf{s}}\geq 1/m$ and $p_{k+1,\mathbf{s}}< 1/m$ and $k = m$ if $p_{i,\mathbf{s}} = 1/m$ for all $i\in [m]$. Note that such an index $k$ must exist otherwise $\sum_{i\in [m]} p_{i,\mathbf{s}} < 1$ which contradicts the law of total probability. It is clear that Inequality~(\ref{eq:LB_SW_softmax}) holds when $k = m$. For the case $k<m$, we have 
\vspace{-1pt}
\begin{align*}
SW(\mathbf{s}) &= \sum_{i\in [m]} p_{i,\mathbf{s}}\cdot u(s_i)\\
&= \sum_{i=1}^k \Big(p_{i,\mathbf{s}} - \frac{1}{m}\Big) u(s_i) + \frac{1}{m}\sum_{i=1}^m u(s_i)
 + \sum_{i=k+1}^m \Big(p_{i,\mathbf{s}} - \frac{1}{m}\Big) u(s_i)\\
&\geq \sum_{i=1}^k \Big(p_{i,\mathbf{s}} - \frac{1}{m}\Big) u(s_k) + \frac{1}{m}\sum_{i=1}^m u(s_i)
 + \sum_{i=k+1}^m \Big(p_{i,\mathbf{s}} - \frac{1}{m}\Big) u(s_k)\\
&=\Big(\sum_{i=1}^m p_{i,\mathbf{s}} - 1\Big)u(s_k) + \frac{1}{m}\sum_{i=1}^m u(s_i)\\
&= \frac{1}{m}\sum_{i=1}^m u(s_i).
\end{align*} 
\vspace{-0pt}
Hence the inequality~(\ref{eq:LB_SW_softmax}) is valid. 

Now, we are ready for Theorem~\ref{thm:PoA_SF} and its proof. 

\begin{thm}\label{thm:PoA_SF}
The PoA of the egoistic election game using any cross-monotone and order-preserving WP function is upper bounded by~$m$. 
\end{thm}
\begin{proof}
Let $\mathbf{s} = (s_i)_{i\in [m]}$ be a PSNE and $\mathbf{s}^* = (s_i^*)_{i\in [m]}$ be the optimal profile. 
Let $\ell = \argmax_{i\in [m]} u(s_i^*)$ be the index of the party players with the maximum social utility with respect to~$\mathbf{s}^*$. Recall that $m\cdot SW(\mathbf{s})\geq \sum_{i\in [m]}u(s_i)$. 
Our goal is to prove Inequality~(\ref{eq:thm4c}):  
\begin{equation}\label{eq:thm4c}
\sum_{i\in [m]}u(s_i)\geq u(s_{\ell}^*).
\end{equation}
By Lemma~\ref{lem:dominated_Not_PSNE} we know that each $i\in [m]$, strategy $s_i$ is not surpassed by $s_i^*$ since $\mathbf{s}$ is a PSNE. Therefore, it suffices to consider the following cases.  
\begin{enumerate}[label=(\alph*)]
    \item For all $i\in [m]$, $s_i\leq s_i^*$. In this case, Inequality~(\ref{eq:thm4c}) holds since
   \begin{enumerate}[label=(\roman*)]
    \item $u_{\ell}(s_{\ell})\geq u_{\ell}(s_{\ell}^*)$,
    \item $u_{j}(s_{j})\geq u_{j}(s_{\ell}^*)$ for each $j\in[m]$. 
    \end{enumerate}
    \vspace{5pt}
    \item For all $i\in [m]$, $u(s_i^*)\leq u(s_i)$. In this case, we have $\sum_{i\in [m]} u(s_i) - u(s_{\ell}^*)\geq \sum_{i\in[m]\setminus\{\ell\}} u(s_i)\geq 0$. Hence, Inequality~(\ref{eq:thm4c}) holds.
    
    \item Suppose that there exists a subset $W\subset [m]$ such that $s_i\leq s_i^*$ for all $i\in W$ and $u(s_j^*)\leq u(s_j)$ for each $j\in \overline{W} := [m]\setminus W$. 
    
    \begin{enumerate}[label=(\roman*)]
    \item Assume that $\ell\in W$. 
    By the arguments similar to (a), Inequality~(\ref{eq:thm4c}) follows from~$u_{\ell}(s_{\ell})\geq u_{\ell}(s_{\ell}^*)$ (by the assumption that $\ell\in W$ and the egoistic property), and $u_{j}(s_{j})\geq u_{j}(s_{\ell}^*)$ for each $j\in[m]\setminus\{\ell\}$ (by the egoistic property).
     \vspace{2pt}
    \item Assume that $\ell\in \overline{W}$. Similar to (b), we have 
        \vspace{-0pt}
        \begin{equation*}
    	\sum_{i\in [m]} u(s_i) - u(s_{\ell}^*)\geq \sum_{i\in \overline{W}} u(s_i) - u(s_{\ell}^*)\geq \sum_{i\in\overline{W}\setminus\{\ell\}}  u(s_i)\geq 0.
    	\end{equation*}
        \end{enumerate}
\end{enumerate}
Together with Inequality~(\ref{eq:LB_SW_softmax}) and (\ref{eq:UB_SW_softmax}), we derive that $SW(\mathbf{s})\geq SW(\mathbf{s}^*)/m$. 
Therefore, we conclude that the PoA is at most~$m$. 
\end{proof}

\paragraph{Remark} In~\cite{LLC2021}, the lower bound on the PoA of the egoistic two-party election game using the softmax WP function is~2. 
Here we further point out that the PoA upper bound is tight when the hardmax WP function is applied. Consider the game instance shown in Table~\ref{tab:AlwaysPSNE_HM_2party}. We can see that the PoA of the game approaches~$m$. Through these examples we know that our PoA bound is tight for $m=2$ using either the hardmax or the softmax function as the WP function, and is also tight for general $m\geq 2$ when the hardmax WP function is applied. 

\paragraph{Remark} In Table~\ref{tab:AlwaysPSNE_HM_2party}, the instance serves as a lower bound approaching~$m$ on the PoS as well since all PSNE share the same social welfare value. Profile $(1,1,\ldots, 1)$ is a PSNE, and it has the social welfare~$\beta/m+3\epsilon$, which approaches $1/m$ of that of the optimal profile, $\beta$, when $\epsilon$ is close to~0.

\begin{table}[ht]
\centering
\caption{An egoistic election game instance of $m\geq 2$ groups that admits a PSNE (hardmax WP; $\beta> 0, n_i=2$ for $i\in\{1,2\}$, $0<\epsilon\ll 1$). The hardmax function is used as the monotone WP function.}\label{tab:AlwaysPSNE_HM_2party}
\begin{tabular}[c]{ c c c c |  c c c c |  c  }
\!\!{\footnotesize $u_1(x_{1,i})$}\!\!\!\!\! & {\footnotesize $u_2(x_{1,i})$}\!\!\!\!\! & \!\!\!\!$\cdots$\!\!\!\!\!\!  & {\footnotesize $u_m(x_{1,i})$}\! & 
    \!\!{\footnotesize $u_1(x_{2,i})$}\!\!\!\!\! & {\footnotesize $u_2(x_{2,i})$}\!\!\!\!\! & \!\!\!\!$\cdots$\!\!\!\!\!  & {\footnotesize $u_m(x_{2,i})$}\! & \!\!\!\!$\cdots$\!\!\!\!\!
    \\
\hline
\!\!$\frac{\beta}{m}+3\epsilon$ & 0 & $\cdots$  & 0 & 
        0 & $\frac{\beta}{m}+2\epsilon$ & $\cdots$  & 0 & \; $\cdots$\\
\!\!$\frac{\beta}{m}$ & $\frac{\beta}{m}$ & $\cdots$  & $\frac{\beta}{m}$ & 
        0 & $\frac{\beta}{m}+\epsilon$ & $\cdots$ & 0 & \; $\cdots$\\
\hline
\end{tabular}
\begin{tabular}[c]{ c c c c }
\!\!{\footnotesize $u_1(x_{m,i})$}\!\!\!\!\! & {\footnotesize $u_2(x_{m,i})$}\!\!\!\!\! & $\cdots$\!\!\!\!\!  & {\footnotesize $u_m(x_{m,i})$}\\
\hline
0 & 0 & $\cdots$  & $\frac{\beta}{m}+2\epsilon$\\
0 & 0 & $\cdots$  & $\frac{\beta}{m}+\epsilon$\\
\hline
\end{tabular}
\end{table}

Finally, we show that without the order-preserving assumption, the PoA can be unbounded.  
\begin{prop}\label{prop:poa_without_order_preserving} 
If the WP function is cross-monotone but not order-preserving, then the egoistic election game may have unbounded PoA.
\end{prop}
\begin{proof}
We show that without the order-preserving assumption, the PoA can be unbounded,
even in the egoistic two-group election game ($m=2$) with a cross-monotone WP function. 
Consider two groups $\mathcal{P}_1$ and $\mathcal{P}_2$. 
Let the WP function be defined by $p_{1,\mathbf{s}}=1$ and $p_{2,\mathbf{s}}=0$ for every profile $\mathbf{s}$,
i.e., group~$\mathcal{P}_1$ always wins regardless of the nominated candidates.
This WP function is monotone and cross-monotone clearly holds.  
Assume that group~$\mathcal{P}_1$ has two candidates $a_1$ and $a_2$, and group~$\mathcal{P}_2$ has two candidates~$b_1$ and~$b_2$.
Let the members' utilities be $u_1(a_1)=2, u_2(a_1)=0, u_1(a_2)=1, u_2(a_2) = M, u_1(b_1)=0$, $u_2(b_1)=M+2$, $u_1(b_2)=0, u_2(b_2) = M+3$, where $M$ is an arbitrarily large constant. Hence, the social utilities are $u(a_1)=2, u(a_2) = M+1, u(b_1)=M+2$ and $u(b_2) = M+3$. 
This game instance is egoistic since $\min\{u_1(a_1),u_1(a_2)\} > \max\{u_1(b_1), u_1(b_2)\}$ and $\min\{u_2(b_1), u_2(b_2)\} > \max\{u_2(a_1),u_2(a_2)\}$. 
The WP function is not order-preserving since at the profile $(a_1,b_2)$ we have $u(b_2) > u(a_1)$ but $p_{2,(a_1,b_2)} = 0 < 1 = p_{1,(a_1,b_2)}$. 
Since group~$\mathcal{P}_1$ wins with probability~$1$ under every profile, its payoff equals the utility of its own members.
Therefore, group~$\mathcal{P}_1$ strictly prefers nominating $a_1$ to~$a_2$ because $u_1(a_1)=2 > 1=u_1(a_2)$, and the profile $(a_1,b_1)$ 
is a PSNE. However, the socially optimal profile is $(a_2,b_1)$ or $(a_2,b_2)$ since $u(a_2)=M+1$ that can be arbitrarily large.
Consequently,
\[
SW((a_1,b_1))=u(a_1)=2,\;\; SW((a_2,b_1))=u(a_2)=M+1,
\]
and the PoA satisfies
\[
\text{PoA} \;\geq\; \frac{SW((a_2,b_1))}{SW((a_1,b_1))} \;=\; \frac{M+1}{2},
\]
which is unbounded as $M\to\infty$.
\end{proof}

\section{Conclusions and Future Work}
\label{sec:future}

From the perspectives of PSNE existence and the PoA, unlike the case of two groups, we have learned that the election game is ``bad'' when more than two groups are involved in the sense that the PSNE is no longer guaranteed to exist and the PoA can be proportional to the number of competing groups even with a cross-monotone and order-preserving WP function. 

We prove that to determine if the election game has a PSNE is {\sf NP}-complete, even for the egoistic election game when the WP function is wired as part of the input. It will be interesting to know if the problem is still {\sf NP}-complete when the WP function is fixed to be the softmax function. 
In spite of the {\sf NP}-completeness result, two parameters of the election game, that is, the number of irresolute groups and the nominating depth, are extracted in this work and utilized to devise an efficient parameterized algorithm for computing a PSNE of the egoistic election game. 
In addition, even when PSNE does not exist, our {\sf FPT} algorithm enumerates a restricted candidate set determined by~$k$ and~$d$,
which may serve as a useful search space for other solution concepts, such as mixed-strategy Nash equilibria.

We have shown that, under a cross-monotone and order-preserving WP function, the PoA of the egoistic election game is upper bounded by the number of groups~$m$, and this also improves the previous bound in~\cite{LLC2021} using the softmax WP function for the two-party case. Our PoA bound is tight for the egoistic two-party election game using either the hardmax or the softmax function as the WP function, and is also tight for general $m\geq 2$ when the hardmax function is applied as the WP function. It will be interesting to know whether this bound is tight for all cross-monotone and order-preserving WP functions. Nevertheless, PoA provides a worst-case measure of inefficiency of a game, so it is also interesting to know if an election game of more group players is still efficient \emph{in average} over all game instances.

An interesting direction is to extend our framework to allow \emph{coalition formation} among groups. 
In such a variant, a coalition may coordinate its nominations and/or share the resulting payoff, which can substantially change the equilibrium structure.
A full treatment requires a precise coalition model (e.g., how coalitions form and what is elected), a principled payoff aggregation rule within a coalition, and an analysis of how these choices affect PSNE existence and its computational complexity under fixed WP rules.
We leave this investigation to future work.


\section*{Acknowledgments}

This work is supported by the National Science and Technology Council of Taiwan under grant no. NSTC 111-2410-H-A49-022-MY2 and NSTC 112-2221-E-032-018-MY3.

\section*{CRediT authorship contribution statement}

C.C.L. and C.J.L. proposed the organizational election game concept and formulated the game model. C.C.L. performed the theoretical analysis, including the {\sf NP}-completeness proof, {\sf FPT} algorithm, and the PoA bounds. C.C.L. wrote the original draft. P.A.C. and C.C.L. verified the technical proofs and finalized the manuscript. C.C.H. proposed the voting simulation experiments and the analysis of the experimental results. All authors reviewed the manuscript.

\section*{Declaration of generative AI and AI-assisted technologies in the writing process}

\noindent During the preparation of this work the author(s) used ChatGPT 5.2 in order to polish the writing and readability. After using this tool/service, the author(s) reviewed and edited the content as needed and take(s) full responsibility for the content of the publication.



\bibliographystyle{elsarticle-num}
\bibliography{election_game_2025}

@InProceedings{AJK2014,
    author  = "N. Ailon and T. Joachims and and Z. Karnin",
    title   = "Reducing dueling bandits to cardinal bandit",
    booktitle	= "Proceedings of the 31st International Conference on Machine Learning (ICML'14)",
    year    = "2014",
    pages   = "II-856--864"
}

@InProceedings{ABP2015,
    author  = "E. Anshelevich and O. Bhardwaj and J. Postl",
    title   = "Approximating optimal social choice under metric preferences",
    booktitle = "Proceedings of the 30th AAAI Conference on Artificial Intelligence (AAAI'15)",
    year    = "2015",
    pages   = "777--783"
}

@InProceedings{AP2016,
    author  = "E. Anshelevich and J. Postl",
    title   = "Randomized social choice functions under metric preferences",
    booktitle = "Proceedings of the 31st International Joint Conference on Artificial Intelligence (IJCAI'16)",
    year    = "2016",
    pages   = "46--59"
}

@article{Bra54,
    author  = "R. A. Bradley and M. E. Terry",
    title   = "Rank Analysis of Incomplete Block Designs: I. The Method of Paired Comparisons",
    journal = "Biometrika",
    volume  = "39",
    number  = "3/4",
    year    = "1954",
    pages   = "324--345",
    doi     = "10.2307/2334029"
}

@article{Cal05,
    author  = "S. Callander",
    title   = "Electoral competition in heterogeneous districts",
    journal = "Journal of Political Economy",
    volume  = "113",
    number  = "5",
    year    = "2005",
    pages   = "1116--1145",
    doi     = "10.1086/444405"
}

@article{CW07,
    author  = "S. Callander and C. Wilson",
    title   = "Turnout, polarization, and {D}uverger's law",
    journal = "Journal of Politics",
    volume  = "69",
    number  = "4",
    year    = "2007",
    pages   = "1047--1056",
    doi     = "10.1111/j.1468-2508.2007.00606.x"
}

@article{CP2011,
    author  = "I. Caragiannis and A. D. Procaccia",
    title   = "Voting almost maximizes social welfare despite limited communication",
    journal = "Artificial Intelligence",
    volume  = "175",
    number  = "9--10",
    year    = "2011",
    pages   = "1655--1671"
}

@InProceedings{CDK2017,
    author  = "Y. Cheng and S. Dughmi and D. Kempe",
    title   = "Of the people: voting is more effective with representative candidates",
    booktitle = "Proceedings of the 18th ACM conference on Economics and Computation (EC'17)",
    year    = "2017",
    pages   = "305--322"
}

@article{Del2013,
    author  = "A. Dellis",
    title   = "The two-party system under alternative voting procedures",
    journal = "Social Choice and Welfare",
    volume  = "40",
    number  = "",
    year    = "2013",
    pages   = "263--284",
    doi     = "10.1007/s00355-011-0597-3"
}

@book{Dow57,
  title		= "An Economic Theory of Democracy",
  author      =  "A. Downs", 
  publisher	= "Harper and Row, New York",
  year		= "1957"
}

@incollection{Dug16,
    title		= "Candidate objectives and electoral equilibrium",
    author      = "J. Duggan", 
    editor      = "B. Weingast and D. Wittman",
    booktitle   = "The Oxford handbook of political economy", 
    publisher	= "Oxford University Press",
    year		= "2008",
    pages       = "64--83", 
    doi         = "10.1093/oxfordhb/9780199548477.003.0004"
}

@book{Duv54,
    title       = "Political Parties: their organization and activity in the modern state", 
    author      = "M. Duverger", 
    address     = "London and New York",
    publisher	= "Methuen \& Co. and John Wiley \& Sons", 
    year        = "1954"
}

@article{Fed92,
    author  = "T. Feddersen",
    title   = "A voting model implying {D}uverger's law and positive turnout",
    journal = "American Journal of Political Science",
    volume  = "36",
    number  = "4",
    year    = "1992",
    pages   = "938--962",
    doi     = "10.2307/2111355"
}

@article{Fey97,
    author  = "M. Fey",
    title   = "Stability and coordination in {D}uverger's law: a formal model of preelection polls and strategic voting",
    journal = "American Political Science Review",
    volume  = "91",
    number  = "1",
    year    = "1997",
    pages   = "135--147",
    doi     = "10.2307/2952264"
}

@article{GGS05,
    author  = "G. Gottlob and G. Greco and F. Scarcello",
    title   = "Pure {N}ash equilibria: hard and easy games",
    journal = "Journal of Artificial Intelligence Research",
    volume  = "24",
    number  = "1",
    year    = "2005",
    pages   = "357--406",
    doi     = "10.1613/jair.1683"
}

@article{Hot29,
    author  = "H. Hotelling",
    title   = "Stability in competition",
    journal = "The Economic Journal",
    volume  = "39",
    number  = "153",
    year    = "1929",
    pages   = "41--57",
    doi     = "10.2307/2224214"
}

@InProceedings{KM2015,
    author  = "J. Kulkarni and V. Mirrokni",
    title   = "Robust Price of Anarchy Bounds via LP and Fenchel Duality",
    booktitle = "Proceedings of the twenty-sixth annual ACM-SIAM symposium on Discrete algorithms (SODA'15)",
    year    = "2015",
    pages   = "1030--1049"
}

@article{Kul00,
    author  = "V. Kuleshov and D. Precup",
    title   = "Algorithms for the multi-armed bandit problems",
    year    = "2014",
    journal = "arXiv preprint arXiv:1402.6028",
    doi     = "10.48550/arXiv.1402.6028"
}

@article{MW93,
    author  = "R. Myerson and R. Weber",
    title   = "A theory of voting equilibria",
    journal = "American Political Science Review",
    volume  = "87",
    number  = "1",
    year    = "1993",
    pages   = "102--114",
    doi     = "10.2307/2938959"
}

@article{Pal84,
    author  = "T. R. Palfrey",
    title   = "Spatial equilibrium with entry",
    journal = "The Review of Economic Studies",
    volume  = "51",
    number  = "1",
    year    = "1984",
    pages   = "139--156",
    doi     = "10.2307/2297710"
}

@incollection{Pal89,
    title		= "A mathematical proof of {D}uverger's law",
    author      = "T. R. Palfrey", 
    editor      = "B. Weingast and D. Wittman",
    booktitle   = "Models of strategic choice in politics", 
    publisher	= "University of Michigan Press, Ann Arbor",
    year		= "1989", 
    pages       = "69--91",
    doi         = "10.7907/n9fr9-jvy72"
}

@InProceedings{PR2006,
    author  = "A. D. Procaccia and J. S. Rosenschein",
    title   = "The distortion of cardinal preferences in voting",
    booktitle = "Proceedings of the 10th International Workshop on Cooperative Information Agents (CIA'06)",
    year    = "2006",
    pages   = "317--331"
}

@InProceedings{R2009,
    author  = "T. Roughgarden",
    title   = "Intrinsic Robustness of the Price of Anarchy",
    booktitle = "Proceedings of the 41st Annual Symposium on Theory of Computing (STOC'09)",
    year    = "2009",
    pages   = "513--522"
}

@book{SB98,
    title		= "Reinforcement Learning: An Introduction. Second Edition",
    author      = "R. S. Sutton and A. G. Barto", 
    publisher	= "MIT Press",
    address     =   "",
    year		= "2018"
}

@article{Web92,
    author  = "S. Weber",
    title   = "On hierarchical spatial competition",
    journal = "Review of Economic Studies",
    volume  = "59",
    number  = "2",
    year    = "1992",
    pages   = "407--425",
    doi     = "10.2307/2297961"
}

@article{YBKJ2012,
    author  = "Y. Yue and J. Broder and R. Kleinberg and T. Joachims",
    title   = "The $k$-armed dueling bandits problem",
    journal = "Journal of Computer and System Sciences",
    volume  = "78",
    number  = "5",
    year    = "2012",
    pages   = "1538--1556",
    doi     = "10.1016/j.jcss.2011.12.028"
}

@article{LLC2021,
    author  = "C.-C. Lin and C.-J. Lu and P.-A. Chen",
    title   = "How good is a two-party election game?",
    journal = "Theoretical Computer Science",
    volume  = "871",
    number  = "",
    year    = "2021",
    pages   = "79--93",
    doi     = "10.1016/j.tcs.2021.04.013"
}

@book{DF13,
    title = "Fundamentals of Parameterized Complexity",
    author = "R. G. Downey and M. R. Fellows", 
    publisher = "Springer Publishing Company, Incorporated",
    isbn = "1447155580", 
    year = "2013",
    DOI = "10.1007/978-1-4471-5559-1"
}

@book{Nie06,
    title		= "Invitation to Fixed-Parameter Algorithms",
    author      = "R. Niedermeier", 
    publisher	= "Oxford University Press",
    isbn        = "9780191524158",
    series      = "Oxford Lecture Series in Mathematics and Its Applications",
    year		= "2006", 
    DOI         = "10.1093/acprof:oso/9780198566076.001.0001"
}

@book{FG06,
    title		= "Parameterized Complexity Theory",
    author      = "J. Flum and M. Grohe", 
    publisher	= "Springer, New York",
    year		= "2006", 
    DOI         = "10.1007/3-540-29953-X" 
}

@article{AGS2011,
    title = "Equilibria problems on games: Complexity versus succinctness",
    author  = "C. \`{A}lvarez and J. Gabarro and M. Serna",
    journal = "Journal of Computer and System Sciences",
    volume  = "77",
    number  = "6",
    pages   = "1172--1197",
    year    = "2011", 
    doi     = "10.1016/j.jcss.2011.01.001"
}

@book{OR94,
  title		= "A Course in Game Theory",
  author      = "M. Osborne and A. Rubinstein", 
  publisher	= "MIT Press",
  year		= "1994"
}

@article{nash_1950,
    author  = "J. F. Nash",
    title   = "Equilibrium points in $n$-person games",
    journal = "Proceedings of the National Academy of Sciences",
    volume  = "36",
    number  = "1",
    year    = "1950",
    pages   = "48--49",
    doi     = "10.1073/pnas.36.1.48"
}

@article{nash_1951,
    author  = "J. F. Nash",
    title   = "Noncooperative games",
    journal = "Annals of Mathematics",
    volume  = "54",
    year    = "1951",
    pages   = "289--295",
    doi     = "10.2307/1969529"
}

@InProceedings{DWH2020,
    author  = "Y. Du and S. Wang and L. Huang",
    title   = "Dueling Bandits: From Two-dueling to Multi-dueling",
    booktitle = "Proceedings of the 19th International Conference on Autonomous Agents and MultiAgent Systems (AAMAS'20)", 
    address = "Auckland, New Zealand", 
    publisher = "International Foundation for Autonomous Agents and Multiagent Systems", 
    year    = "2020",
    pages   = "348--356"
}

@InProceedings{DL2014, 
    title = "On Computing Optimal Strategies in Open List Proportional Representation: The Two Parties Case", 
    author = "N. Ding and F. Lin", 
    booktitle = "Proceedings of the 28th AAAI Conference on Artificial Intelligence (AAAI'14)", 
    address = "Paris, France", 
    publisher = "AAAI Press", 
    DOI = "10.1609/aaai.v28i1.8888", 
    year = "2014", 
    month = "June", 
    pages = "1419--1425"
}

@inproceedings{SORR17,
    author = "I. Sabato and S. Obraztsova and Z. Rabinovich and J. S. Rosenschein",
    title = "Real Candidacy Games: A New Model for Strategic Candidacy",
    year = "2017",
    publisher = "International Foundation for Autonomous Agents and Multiagent Systems",
    booktitle = "Proceedings of the 16th Conference on Autonomous Agents and MultiAgent Systems (AAMAS'17)",
    pages = "867--875",
    numpages = "9",
    location = "S\~{a}o Paulo, Brazil", 
    address = "Richland, SC"
}

@inproceedings{HLST21,
    author = "P. Harrenstein and G. Lisowski and R. Sridharan and P. Turrini",
    title = "A Hotelling-Downs Framework for Party Nominees",
    year = "2021",
    isbn = "9781450383073",
    publisher = "International Foundation for Autonomous Agents and Multiagent Systems",
    address = {Richland, SC},
    booktitle = "Proceedings of the 20th International Conference on Autonomous Agents and MultiAgent Systems (AAMAS'21)",
    pages = "593--601",
    numpages = "9",
    location = "Virtual Event, United Kingdom"
}

@article{L00a,
    author  = "J. F. Laslier",
    title   = "Interpretation of electoral mixed strategies",
    journal = "Social Choice and Welfare",
    volume  = "17",
    year    = "2000",
    pages   = "283--292",
    DOI = "10.1007/s003550050021"
}

@article{L00b,
    author  = "J. F. Laslier",
    title   = "Aggregation of preferences with a variable set of alternatives",
    journal = "Social Choice and Welfare",
    volume  = "17",
    year    = "2000",
    pages   = "269--282", 
    DOI     = "10.1007/s003550050020"
}

@article{BFM18,
    title = "Opinion formation games with dynamic social influences",
    author = "V. Bil\`{o} and A. Fanelli and L. Moscardelli",
    journal = "Theoretical Computer Science",
    volume = "746",
    pages = "73--87",
    year = "2018",
    doi = "10.1016/j.tcs.2018.06.025",
}

@InProceedings{AFS19,
    author = "V. Auletta and D. Ferraioli and V. Savarese",
    editor = "M. Alviano and G. Greco and F. Scarcello",
    title = "Manipulating an Election in Social Networks Through Edge Addition",
    booktitle = "Advances in Artificial Intelligence (AI*IA 2019)",
    year = "2019",
    publisher = "Springer International Publishing",
    address = "Cham",
    pages = "495--510",
}

@article{cechlarova2023hardness,
    title = "Hardness of candidate nomination",
    author = "Cechl{\'a}rov{\'a}, K. and Lesca, J. and Trellov{\'a}, D. and Han{\v{c}}ov{\'a}, M. and Han{\v{c}}, Jozef",
    journal = "Autonomous Agents and Multi-Agent Systems",
    volume = "37",
    number = "2",
    pages = "37",
    year = "2023",
    publisher = "Springer",
    doi = "10.1007/s10458-023-09622-9"
}

@inproceedings{LLC2024,
    author    = "Lin, C.-C. and Lu, C.-J. and Chen, P.-A.",
    title     = "How Bad Can an Election Game of Two or More Parties Be?",
    booktitle = "The 6th Games, Agents, and Incentives Workshop (GAIW-24), AAMAS 2024 Workshops",
    address   = "Auckland, New Zealand",
    month     = "May",
    year      = "2024",
    pages     = ""
}

@book{OR1994CourseGT,
    author    = "Osborne, M. J. and Rubinstein, A.",
    title     = "A Course in Game Theory",
    publisher = "The MIT Press",
    address   = "Cambridge, MA",
    year      = "1994",
    isbn      = "9780262650403"
}

@inproceedings{CS2005IteratedDominance,
    author    = "Conitzer, V. and Sandholm, T.",
    title     = "Complexity of (Iterated) Dominance",
    booktitle = "Proceedings of the 6th ACM Conference on Electronic Commerce (EC '05)",
    year      = "2005",
    pages     = "88--97",
    publisher = "ACM",
    address   = "New York, NY, USA",
    doi       = "10.1145/1064009.1064019"
}

@article{GKZ1993EliminatingDominated,
    author  = "Gilboa, I. and Kalai, E. and Zemel, E.",
    title   = "The Complexity of Eliminating Dominated Strategies",
    journal = "Mathematics of Operations Research",
    year    = "1993",
    volume  = "18",
    number  = "3",
    pages   = "553--565",
    doi     = "10.1287/moor.18.3.553"
}

\end{document}